\documentclass[bibliography=totocnumbered,a4paper]{scrreprt}

\usepackage[UKenglish]{babel}
\usepackage[T1]{fontenc}
\usepackage[ansinew]{inputenc}
\usepackage{csquotes}

\usepackage{graphicx}
\usepackage[long]{datetime}
\usepackage{tabularx}
\usepackage{paralist}
\usepackage[section]{placeins}
\usepackage{alltt}
\usepackage{subfigure}
\usepackage{enumitem}
\usepackage{changepage}
\usepackage{color}
\usepackage{etoolbox}
\usepackage{calc}
\usepackage{algpseudocode}
\usepackage{amsmath}
\usepackage{algorithm}
\usepackage{msc}

\newif\iftech

\expandafter\ifstrequal\expandafter{\jobname}{techrep}{\techtrue}{\techfalse}

\newcommand{\techhds}[2]{\iftech{#1}\else{#2}\fi}
\newcommand{\tech}[1]{\techhds{#1}{}}
\newcommand{\hds}[1]{\techhds{}{#1}}
\newcommand{\thisdoc}{\techhds{report}{HDS}}

\newcommand{\VTNote}[1]{}
\newcommand{\SASNote}[1]{}

\newcommand{\enc}[3]{\mathit{Enc}_{#1}(#2;#3)}
\newcommand{\renc}[2]{\mathit{ReEnc}(#1;#2)}
\newcommand{\dec}[2]{\mathit{Dec}_{#1}(#2)}
\newcommand{\aes}[2]{\mathit{SymmEnc}_{#1}(#2)}
\newcommand{\aesdec}[2]{\mathit{SymmDec}_{#1}(#2)}
\newcommand{\CRT}{\mathit{CRT}}
\newcommand{\RT}{\mathit{RT}}
\newcommand{\SHA}[1]{h(#1)}
\newcommand{\com}[1]{c(#1)}
\newcommand{\epk}{\mathit{PK}_E}
\newcommand{\ppk}{\mathit{PK}_P}
\newcommand{\psk}{\mathit{SK}_P}
\newcommand{\kparam}{k}

\newcommand{\PaV}{Pr\^{e}t \`{a} Voter}
\newcommand{\commentOut}[1]{}
\newcommand{\commentOutForJETS}[1]{}

\newcommand{\VTcomment}[1]{}
\newcommand{\SAScomment}[1]{}

\newcommand{\cand}{\mbox{\it cand}}
\newcommand{\BC}{b}
\newcommand{\SC}{G}
\newcommand{\Mix}{\mbox{\it RGen}}  
\newtheorem{claim}{Claim}

\newcommand{\emphbox}[1]{\begin{center}\framebox{\parbox{0.93\textwidth}{\bf #1}}\end{center}}

\makeatletter
    \def\thebibliography#1{\chapter*{References\@mkboth
      {REFERENCES}{REFERENCES}}
      \addcontentsline{toc}{chapter}{References}
      \list
      {[\arabic{enumi}]}{\settowidth\labelwidth{[#1]}\leftmargin\labelwidth
	\advance\leftmargin\labelsep
	\usecounter{enumi}}
	\def\newblock{\hskip .11em plus .33em minus .07em}
	\sloppy\clubpenalty4000\widowpenalty4000
	\sfcode`\.=1000\relax}
    \makeatother

\def\doccontext{vVote}
\def\doctitle{Technical Report}
\def\docauthor{\ \\ Chris Culnane, Peter Y A Ryan, \\ Steve Schneider and Vanessa Teague}

\def\docversion{4.0}
\def\docdate{\date}

\usepackage[bookmarks=true]{hyperref}
\hypersetup{
    pdftitle={\doctitle}, 
    pdfauthor={\docauthor}, 
    pdfsubject={\doctitle}, 
    pdfkeywords={\doccontext, \doctitle}, 
    colorlinks=true, 
    linkcolor=blue, 
    citecolor=black, 
    filecolor=black, 
    urlcolor=green, 
    linktoc=page 
}%

\makeatletter
\renewcommand*\bib@heading{%
  \section{\refname}\label{sec:references}%
  \@mkboth{\sectionmarkformat \refname}{\sectionmarkformat \refname}%
  }
\makeatletter

\newlength{\descriptionindent}
\newcommand{\request}[1]{\hyperref[request-#1]{\bf(request #1)}}

\newcommand{\sequencefigure}[1]{\begin{adjustwidth}{-1in}{-1in}\centering\includegraphics[scale=0.7]{#1}\end{adjustwidth}}

\newcommand{\pret}{Pr\^{e}t \`{a} Voter}

\begin{document}

\begin{titlepage}

\flushright
\rule{16cm}{5pt}\vskip1cm
{\Huge vVote: a Verifiable Voting System}\\
\vspace{1.5cm}

\vspace{4cm}
 \LARGE{ 
Technical Report Version \docversion \\}
\vspace{1.5cm}
\LARGE{\docdate \\}
\vspace{1.5cm}
\docauthor\\
\vfill
\rule{16cm}{5pt}

\end{titlepage}

\tableofcontents


\chapter*{Abstract}

\addcontentsline{toc}{chapter}{Abstract}


  The \PaV{} cryptographic voting system was designed to be
  flexible and to offer voters a familiar and easy voting
  experience. In this paper we present a case study of our efforts to
  adapt \PaV{} to the idiosyncrasies of elections in the
  Australian state of Victoria.  This technical report includes
  general background, user experience and details of the cryptographic protocols and human processes.
  We explain the problems, present solutions, then analyse their
  security properties and explain how they tie in to other design
  decisions.  We hope this will be an interesting case study on the
  application of end-to-end verifiable voting protocols to real
  elections.  

The prior version of this Technical Report was published on arXiv on the 17th September 2014.  This version has been updated with an
account of the election run.  This is the extended version of our paper \cite{DBLP:journals/tissec/CulnaneRST15}.  The source code is available at 
\url{https://bitbucket.org/vvote/}

The team involved in developing the vVote design described in this report were: Craig Burton, Chris Culnane, James Heather, Rui Joaquim, Peter Y. A. Ryan, Steve Schneider and Vanessa Teague.


\chapter{Introduction}\label{sec:introduction}
The potential advantages of electronic voting are obvious but the risks are not.  
The challenge is to obtain the obvious advantages for convenience and accessibility while preserving reasonable vote privacy and producing a rigorous evidence trail that will stand up to dispute.

This paper details a design for end-to-end verifiable voting in the Australian state of Victoria, based on the \pret{} end-to-end verifiable voting system \cite{RyanBHSX09}.   The system ran successfully in the state election in Victoria (Australia) in November 2014, taking a total of 1121 votes from supervised polling places inside Victoria and overseas.  
The proposed protocol is end-to-end verifiable, meaning that there are no 
 human or electronic components which must be trusted for guaranteeing the integrity of the votes (although vision 
impaired voters must assume that at least one device reads accurately to them). 
There are probabilistic assumptions about the number of voters who confirm correct printing of some 
\PaV{} ballots, the number who check that their printout matches their intended 
vote, and the number who check that their receipt appears on the Web Bulletin Board (WBB).  It also provides 
voters with evidence of malfeasance, assuming that they check the signature on 
their receipt before they leave the polling station.
Since this is a polling-station scheme, we do not address eligibility
verifiability.  Prevention of ballot stuffing is by existing procedural
mechanisms.

\section{End-to-end verifiability}
End-to-end verifiability consists of three pieces of evidence:
\begin{description}
\item[Cast-as-intended verification: ] Each voter gets evidence that their
vote is cast as they intended,
\item[Counted-as-cast verification: ] Each voter gets evidence that their vote is included unaltered in the tally,
\item[Universally verifiable tallying: ] Everyone can check that the
list of (encrypted) cast votes produces the announced election outcome.
\end{description}

This project does not achieve verifiability all the way to the announcement of the election result, because it runs alongside an existing paper-based system that does not provide a universally verifiable tallying proof.   Nevertheless we believe there are significant advantages to using an end-to-end verifiable electronic voting system for part of an election.   See \ref{sec:vicChallenges} for more details.  In summary, the vVote system provides

\begin{itemize}
\item cast-as-intended verification,
\item counted-as-cast verification and
\item an output list of votes, with a universally verifiable proof.
\end{itemize}

Other forms of electronic voting (or indeed of paper-based voting) 
might provide some, but not all, of these properties.  For example,
some Australian states offer a simple electronic ballot marker, in 
which the voter uses a computer to help them fill in, and then print, a ballot that is put into a ballot box and counted along with the ballots that are completed manually.  This provides cast-as-intended verification for sighted voters, because the voter can look at their printed ballot and check that it reflects their intention.  It does not provide counted-as-cast verification, instead relying on procedures for the secure transport of paper ballots (like the paper system does).  It delegates tallying verification to scrutineers who observe the paper count (if there is one).

The VEC opted for \pret{} over the simpler alternative of a plain electronic ballot marker because it provides for electronic transfer of ballot information from distant supervised locations, supported by verifiable evidence of correctness.  This was regarded as particularly important for distant polling places ({\it e.g.} overseas) and for allowing any voter to vote at any polling place.  Since this project commenced, a problem in the transport of West Australian Senate ballot papers in the 2013 federal election has focused national attention on the security of processes for transporting paper ballots.

\section{\pret{} overview}
\pret{} uses a ballot form that is printed before voting, with a list of candidates printed in a random order, and an encrypted version of the same list.  Voters select or number the candidates by filling in boxes adjacent to the candidate names (in the Victorian protocol, they will have computerised assistance). 
They keep the list of marked boxes and the encrypted candidate list, and shred the human-readable candidate list.    The two main properties of \pret{} are privacy and end-to-end verifiability.  

End-to-end verifiability is achieved in vVote, following the \pret{} structure, as follows:

\begin{description}
\item[Cast-as-intended verification: ] \ \\[-3ex]
	\begin{itemize}
	\item {\bf ballot printing confirmation } Each voter has the opportunity to confirm that the printed candidate lists on some ballots match the encrypted version on the same ballot.
	\item {\bf preference printing confirmation } Each voter writes their own preferences (or in the case of vVote, checks that their own preferences are correctly written) on the other side of the ballot.
	\end{itemize}
\item[Counted-as-cast verification: ] Each voter gets the opportunity to check that their (encrypted) ballot appears in a public list of received votes,
\item[Universally verifiable tallying: ] Everyone can check that the
list of (encrypted) cast votes produces the announced output votes, by downloading and checking a public electronic proof.
\end{description}

\section{Necessary procedures}
Academic papers can simply state the assumptions under which the protocol is secure; practical deployments must ensure that those assumptions are met, or the system may not protect privacy or provide evidence of a
correct election result.  This technical report details the procedures that need to be followed in order for the system to attain the verifiability and privacy properties necessary for state elections.

{\bf End-to-end verification} depends on voters performing proper procedures for verifying that their vote is cast as they intended and included in the count.  All voters must be explicitly encouraged to perform some ballot confirmations on the printed ballots before voting, to check that the computer-assisted preference list matches their intended vote, and to look up their vote later on the public list of accepted votes.  If the announced election outcome is disputed, then the evidence supporting it depends crucially on the number and distribution of these confirming procedures that were actually performed.  For example, if there is one polling place in which voters were not informed of the ballot confirmation procedure, then all of the votes cast at that polling place lack any supporting evidence.  

{\bf Privacy} depends on the immediate destruction of the human-readable candidate list.  Unlike ordinary completed ballots, the human-readable candidate lists also include a unique serial number that matches the one on the voter's receipt.  Hence the candidate lists prove how individuals voted.  Shredding the candidate list must be enforced at all polling places, immediately after the person votes.

\section{The protocol}
End-to-end verifiable election protocols are well studied in the academic 
literature, but until recently have not been deployed in public elections. In 
2011 the Victorian Electoral Commission (VEC) approached the \PaV{} team to
investigate 
adapting the scheme to the special requirements of the Victorian parliamentary 
elections, which use both Instant Runoff Voting (IRV, also called ``alternative vote'' in the UK, and simply ``preferential voting'' in Australia) and 5-seat Single Transferable Vote (STV)\footnote{For 
more information on various election methods, please refer to the appendix in 
\cite{Xia10:VersatilePaV}.}.  The first version of the final system has been recently delivered to the VEC at the time of writing, and a process of testing and review will soon be underway.  This document is intended as an aid for testing, review and security analysis.  The system is expected to be ready for the next Victorian State election in November 2014. \VTNote{Update}

The proposed protocol is universally verifiable, meaning that there are no 
hardware, software, or human components that must be trusted for guaranteeing the integrity of the votes.\footnote{Vision 
impaired voters must assume that at least one device reads accurately to them.} 
There are probabilistic assumptions about the number of voters who confirm correct printing of some 
\PaV{} ballots, the number who check that their printout matches their intended 
vote, and the number who check that their receipt appears on the Web Bulletin Board (WBB).  It also provides 
voters with evidence of malfeasance, assuming that they check the signature on 
their receipt before they leave the polling station.

Since this is a polling-station scheme, we do not address eligibility
verifiability.  Prevention of ballot stuffing is by existing procedural
mechanisms.

The main departure from standard \PaV{} is the use 
of a computer to assist the user in completing the ballot.  This is referred to as
an ``electronic ballot marker'' (EBM).
The EBM is trusted for vote privacy, which is different from standard \PaV{} in 
which the voter does not need to communicate her vote to any (encryption) 
device. This modification is necessary for usability, because a vote can 
consist of a permuted list of about 30 candidates.  It seemed infeasible for a 
voter to fill in a \PaV{} ballot form without assistance.  Indeed, simply 
filling in an ordinary paper ballot with about 30 preferences is a difficult 
task.\footnote{Since some people deliberately vote informally, it is 
difficult to say exactly what percentage of people accidentally disenfranchise themselves by 
incorrectly filling in their vote. About 2\% of votes in the 2006 state election were ruled informal because of ``numbering errors''  \cite{VEC2006} \VTNote{Cite VEC report on 2006 election - SAS: Done}, but the overall informality rate is closer to 10\%, especially when there are lots of candidates on the ballot.  See
{\tt https://www.vec.vic.gov.au/Results/stateby2012distributionMelbourneDistrict.html } for an example.} Computerised assistance is an important 
benefit of the project, and trusting the device for privacy seemed an almost 
unavoidable result of that usability advantage.\footnote{In principle one could
use an EBM to fill in a series of ballots and only cast one of them, without
telling the device which one.  This is too much work for voters.} Hence our scheme
depends on stronger privacy
 assumptions than standard \PaV{}. Providing privacy for complex ballots is 
notoriously difficult, and is further complicated by some details of 
{Victorian} elections that are described below. Our system provides 
privacy and receipt-freeness under reasonable assumptions about the correct 
randomised generation and careful deletion of secret data, and of course 
assuming a threshold of decryption key sharers do not collude. It does not 
fully defend against the ``Italian Attack,'' or all other subtle coercion issues, 
but neither does the current paper-based system.  We make this more precise 
below.

\section{Challenges of Victorian Voting} \label{sec:vicChallenges}

\PaV{} was designed originally for first-past-the post voting, in which
each voter chose a single candidate~\cite{chaum05:e-vote}. Subsequent papers extended the scheme to
more complex types of elections~\cite{ryan106:e-vote,pavPaillier,RyanBHSX09,Xia10:VersatilePaV}.  

The state of {Victoria}, like many
other {Australian} states, runs simultaneous elections for two houses of parliament, the Legislative Assembly (LA) and the Legislative Council (LC), both of which use ranked-choice voting.  Each LA representative is elected by IRV with
compulsory complete preference listing, with rarely more than 10 candidates.  Members of the Legislative Council (LC) are elected in 5-member electorates using STV.  Voters typically choose from among about 30
candidates---they rank at least 5, and up to all
candidates in their order of preference.

Because LC voting is quite complex, voters are offered a shorthand called ``Above the line'' (ATL) voting, which allows them to select a complete preference ranking chosen by their favourite political group (usually a party).  Each political group (of which there are about 12)
registers a (complete) STV vote in advance with the electoral commission.  When someone votes ATL and chooses that group's ticket, this is equivalent to copying out their STV vote.    

Traditionally, both LC voting options are presented on the same ballot
paper.  The ATL group selections are presented on
top of a thick line (hence the name); the
full STV options are shown below the line (and hence called ``below the line'' (BTL) votes).

Each polling place must accept votes from a resident of anywhere in the 
state.   Hence our system must produce \PaV{} ballots for every electoral division in both the LA and the LC,
available at every polling place.  This is a significant challenge for \PaV{},
but \PaV{} confers the great advantage of verifiability on these votes. The existing
methods of verifiable paper counting do not work with 
this requirement.  For the large fraction of people who vote outside their home 
electorate, completed paper ballots must be sent to the home electorate by 
courier, usually arriving after the polling-station count has been completed 
and after observers have departed.   

This system will not be responsible for all of the votes cast in the
upcoming state election, so it will have to combine with existing
procedures for casting and counting ordinary paper ballots.  For LA
and LC-ATL votes this is straightforward. However, LC-BTL votes are
complicated.  Even those cast on paper must be tallied
electronically---in the existing system they are manually entered and
then electronically tallied.  The authorities then make complete
vote data available to allow observers to check the count.\footnote{These
  procedures are also under review and improvement, but are out of the
  scope of this paper.}
This is why the system does not achieve a complete universally verifiable
tally.
The proof that this system produces would be sufficient for end-to-end verifiability if it carried all votes in the election, but it is not
possible to do STV tallying (whether verifiable or not) on a subset
of votes.  As it is, the scrutineers who observe the paper count will have to check that the 
publicly verifiable output from vVote matches the votes that are added to the paper count.  \VTNote{This needs to be added to the procedures list too.}

Preferential elections are vulnerable to coercion through signature
attacks \cite{NOSRC:eprints/hal/Cosmo2007}, commonly referred to as
Italian attacks.
The system proposed here does not address this attack, primarily
because it will work alongside a paper system that is also susceptible
to it.  Our system also reveals whether a person voted ATL or BTL.
This is unlikely to have political consequences.

Another challenge is producing an accessible solution for voters who
cannot fill out a paper ballot unassisted.  This is a primary
justification for the project, but producing a truly verifiable
solution for such voters is extremely difficult, because many of them
cannot perform the crucial check that the printout matches their
intention (though see \cite{DBLP:journals/cryptologia/ChaumHPV09} for
a verifiable and accessible protocol). We provide a way for them to
use any other machine in the polling place to do the check, in which
case the cast-as-intended property depends upon at least one of the
machines in the polling station not colluding with the others to
manipulate the vote.

\section{Specific design choices}
\subsection{Cast-as-intended verification}
Wombat \cite{wombat2},  VoteBox \cite{VoteBox} and several other 
polling-station end-to-end verifiable voting schemes  guarantee integrity by using ``Benaloh 
challenges,'' \cite{BenalohSimple06} which require filling in the vote more than once. 
This would be 
time-consuming for 30-candidate STV. It would perhaps be possible to make 
challenges easier (for example, by letting the device remember the last vote), 
but the integrity guarantees still depend on the voter performing quite a 
subtle randomised protocol.  We have opted for \PaV{}, in which voters may confirm the correctness of the 
unvoted ballot form.  This confirmation process (called ``auditing'' in older versions) can be 
completed with assistance without compromising privacy, because it occurs before the person votes.  It does not require the voter to redo their (possibly quite complicated) vote.  It also provides dispute resolution and some accountability: there is no need to take the voter's word for how they voted.  A ballot confirmation check that completes with an invalid proof can be used as evidence; an attempted ballot 
confirmation check that does not complete at all can have multiple (human) witnesses.

Ballot confirming is separate from voting, so additional
ballot confirming by independent observers would be a 
convenient and practical addition to voter-initiated ballot confirmations.  It would be easy for polling-place observers to see that the confirmation process did not involve casting any votes.  (Wombat, StarVote and some other systems also separate the process of generating an encrypted vote from casting it.)

\commentOut{iii. Lots of different ways you could use this, e.g. authorities guarding against external threats (where the pollworkers know whether they're really going to vote), vs plainclothes auditors who pretend to be voters but then don't vote, etc.  The option for individual voters to audit means that those who distrust the authorities can also check for themselves.
iv.	
v.	There's a bit of subtlety about who should know which ballots are going to be audited, and how hard this makes it to stuff ballots vs print bad candidate lists.
}

These processes are additive in the sense that they do not interfere with each other: the audits and inferences associated with particular trust assumptions are not affected by other audits based on different trust assumptions. 

\subsection{Unified Scanner and EBM}
We have already described why completing the ballot needs to be assisted by a computer.  Our original design \cite{EVOTE2012:VEC} included separate steps for filling in the ballot and then scanning the printed receipt.  This was designed to separate the information of how the person voted from the knowledge of what their receipt looked like: the EBM learnt how the person voted, but could not subsequently recognise their ballot (and hence link it to the individual voter), while the scanner knew the receipt but did not know the corresponding plaintext.  However, user studies at the VEC determined that a three-step voting process was too cumbersome for use.  Also the necessity of print-on-demand meant that there was already an Internet-connected machine in the polling place that was trusted for maintaining privacy of the information on the printed ballot, including which candidate ordering corresponded to which receipt.  For both these reasons, the new protocol now unifies the job of the scanner and the EBM, though it retains a separate print-on-demand step.  The voter first collects their ballot form, and has an opportunity to perform a confirmation check on it, then goes to an EBM to fill in the ballot, then the EBM sends the receipt electronically and also prints a paper record for the voter to check.  This now means there are two online manchines in the polling place (the EBMs and the ballot printers) that are trusted for vote privacy.

\subsection{Print on Demand}
This project necessitated a new protocol for the verifiable printing on demand of \PaV{} ballot forms.  The crucial requirement is that voters (and others) can perform a confirmation check on some ballot forms for correctness \emph{without compromising voter privacy}, because the check occurs before the person votes.  Voters then vote on ballot forms that have not had a confirmation check.  Ballot forms that have had a confirmation check cannot be voted on because their associated ciphertexts have now been decrypted, and privacy would be lost on such votes.

The integrity of \PaV{} depends crucially on proper construction of the printed ballot forms, meaning that the plaintext candidate list that the voter sees must match the encrypted values for that ballot.  This technical report details opportunities for confirming their correct construction and printing.  The construction is very computationally efficient and retains most of the desirable properties of existing print-on-demand proposals in the literature.  The information flow of our scheme is similar to Markpledge 2 \cite{adida2006ballot}, though the confirming is different.  The main idea is that the device encrypts the vote directly using randomness generated by others. The protocol was first presented in JETS \cite{culnaneEtAl2013PoD}.

Also the system must address the question of ``kleptographic'' privacy attacks \cite{DBLP:conf/etrics/GogolewskiKKKLZ06}, in which the (public) ciphertexts contain deliberately poorly-chosen randomness that exposes the vote.  This is possible whenever the entity building the ciphertexts also controls all the randomness used.  This problem is addressed by distributing the process of inputting randomness into ciphertexts.  

This proposal is designed so that the entity that builds the ciphertexts (the printer) has a deterministic algorithm to follow.
This provides a way to distribute the expensive cryptographic operations to the network of printers, whilst retaining the central, distributed, generation of randomness to maintain the following three properties:

\begin{itemize}
\item ensuring the candidate lists are randomly generated,
\item ensuring no single generating entity knows all the (plaintext) candidate lists, and
\item ensuring extra information about the candidate list cannot be leaked in the ballot ciphertexts (as in kleptographic attacks ).
\end{itemize}

We then devise a confirming mechanism to ensure correctness of the printed ballot forms.

\subsection{Randomised Partial Checking}
The exact choice of mixnet is independent of other aspects of the protocol,
but in this implementation we have selected a mixnet based on Randomised Partial Checking \cite{jakobsson02:e-vote}.  This was partly due to efficiency, and partly
to the ease of explaining to the public how the mixnet works.

However, improvements in both the implementation and the efficiency of 
zero-knowledge shuffling proofs \cite{terelius2010proofs, furukawa2001efficient, neff01:e-vote} could make them a reasonable alternative in future versions.  In theory they have superior
properties, because their privacy and soundness are stronger, can be proven formally, and rely on weaker assumptions than those of RPC.  However, they remain computationally intensive and difficult to explain.

\section{Related Work}
In the USA, permanent paper records such as Voter Verified Paper Audit Trails (VVPAT) or opscans are a
common means of achieving software 
independence~\cite{rivest08:e-vote}.  However, this does
not solve the problem of secure custody and transport of the paper
trail.  Furthermore, performing rigorous risk-limiting
audits seems intractable for IRV \cite{Magrino:irv},
let alone for 30-candidate STV.

The most closely related project is the groundbreaking use of
Scantegrity II in binding local government elections in Takoma Park,
MD \cite{SIITakPk}.  Our project has very similar privacy and
verifiability properties.  However, both the overall election size and the complexity
of each ballot are greater for our system.  Although the Scantegrity II scheme appears to have 
been highly successful in the context of the Takoma Park elections, \PaV{} is more appropriate for our
application.  Scantegrity II is inherently for single-candidate
selections.  It has been adapted to IRV in Takoma Park by running a
separate single-candidate election for each preference, but would be
difficult to adapt to 30-candidate preference lists. Even with
computer assistance, a 30 by 30 grid of invisible ink bubbles seems
too complicated for most voters.

The STAR-Vote project proposed for Travis County, TX \cite{starvote-jets} represents an interesting combination of 
end-to-end verification techniques and risk limiting audits.  
STAR-Vote retains a human-readable paper record for auditing purposes alongside the end-to-end verifiable cryptographic data.  Cast-as-intended verification of the end-to-end verifiable part is achieved by a novel interpretation of Benaloh's simple challenge process~\cite{BenalohSimple06}, in which voters can choose
either to cast their ballot into a special ballot box or to spoil it and start again.  We hope our observations might be helpful in the final stages of the STAR-Vote design process.

\section{Prior work and paper overview}
In a previous paper \cite{EVOTE2012:VEC} we gave an overview of this
project, including the context of Victorian voting and some
ideas about how we would implement the protocol.  A followup version \cite{burton2012using}  gave more details and some preliminary security analysis.  The print on demand protocol was presented in \cite{culnaneEtAl2013PoD}.  Here we present all of the protocol, including both the cryptographic protocol and the human procedures to be followed in the polling place
and at the electoral commission.  Our aim is for a comprehensive
analysis of the protocol's security, including the assumptions on
which privacy depends, a precise explanation of the kind of
verifiability achieved, and a clear statement of the issues that
remain.  

An overview including the main system components and an account of the voting experience is contained in Section~\ref{sec:system-overview}, which includes a statement of the main security claims.  Many of the system's security properties depend on proper procedures in the polling place---these are detailed in Section~\ref{sec:polling-place-procedures}. The individual system components are described in Section~\ref{sec:system-component-details}.  Section~\ref{sec:robustness-recovery} contains mechanisms for achieving robustness in the presence of certain failures.  A comprehensive and rigorous threat analysis is contained in Section~\ref{sec:security-analysis}.

\chapter{System Overview}\label{sec:system-overview}

\section{System Components}\label{sec:system-components}
The system has the following main components:

\begin{description}
	\item[Public Web Bulletin Board (Public WBB): ] an authenticated public broadcast channel with memory.
	\item[Private Web Bulletin Board (Private WBB): ] a robust secure database which receives messages, performs basic validity checks, and returns a signature.  Validly signed messages are guaranteed, under certain assumptions, to appear subsequently on the Public WBB. 	
	\item[Print-on-demand printer: ] a combination of a computer and printer which generates \pret{} ballots in advance of the election, then prints them on demand.
	\item[Randomness Generation Service: ] a collection of servers that produce randomness for the print on demand process. 
	\item[Electronic Ballot Marker (EBM): ] a computer that assists the user in filling in a \pret{} ballot.
	\item[Cancel Station: ] a supervised interface for cancelling a vote that has not been properly submitted or has not received a valid Private WBB signature.
	\item[Mixnet: ] a set of (preferably independently managed and hosted) Mix servers that produces a noninteractive, universally verifiable proof of a shuffle and decryption (of encrypted votes) and posts it to the Public WBB.
	\item[Election key sharers: ] authorities who share the key used to decrypt votes at
  the end of the election.
\end{description}

\section{Voter Experience Overview}\label{sec:voter-experience}
This section gives an overview of the verification process, listing all the steps for a voter to verify that their vote is cast as they intended and properly included in the count, and also the process for universal verification of the output. 

\VTNote{A lot of details are deliberately left out here, e.g. procedures for thwarting chain voting, but I think they should go later or we'll all get confused.  e.g. this part says ``the voter shreds the candidate list...'' but doesn't explain how that's enforced.  I think that should go in ``procedural details'' below.}

\begin{figure}[th]
	\centering
		\includegraphics[width=0.95\textwidth]{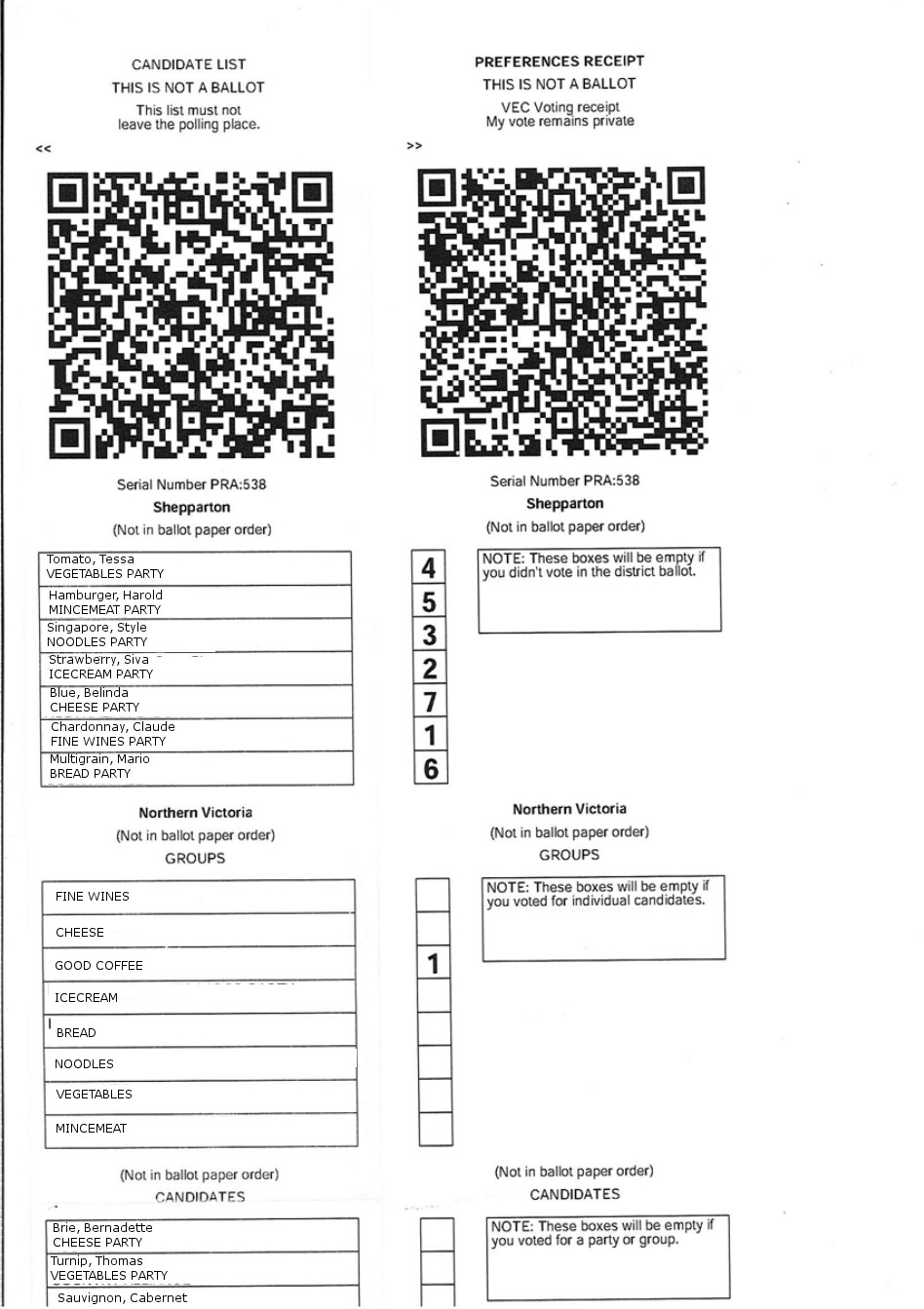}
	\caption{Separate vote printouts (truncated at the bottom).  The voter collects the left side (the candidate list) when they are marked off.  The EBM then prints the right side with the voter's preferences.  Note the matching serial numbers.}
	\label{fig:ballot}
\end{figure}

Recall that voters cast an IRV vote for a Legislative Assembly district and then for their Legislative Council region either a full STV vote or an 'ATL' shorthand.  (See Section~\ref{sec:vicChallenges}).  
A printed ballot therefore consists of:
\begin{itemize}
\item A human-readable serial number (shortened to SerialNumber below),
\item a human-readable district name (which also determines the region),
\item a human-readable randomly ordered list of the candidate names for the LA district,
\item a human-readable randomly ordered list of the candidate names for the LC region,
\item a human-readable list of group names (for LC-ATL voting),  
\item a QR code containing all this data, plus a WBB digital signature on it.
\end{itemize}

Three races: a named district race and the two exclusively votable parts of the region race (a group-list vote or voting for individual candidates)
A list of candidate names for the District and candidate voting part of the region race
A list of group names for the group-list part of the region race

Figure~\ref{fig:ballot} shows the ballot form on the left side, and also
the preferences as printed out by the EBM.  The process for receiving the ballot form, then casting a vote on it while verifying that the vote matches the voter's intention, is as follows:

\begin{description}
\item[Pollworker:] authenticates the voters (using whatever method is traditional) and sends a print request to the Print On Demand device specifying the district/region they can vote in,\footnote{More generally, it is the pollworkers' responsibility to authenticate the voter, request the ballot(s) that the person is eligible to vote on, and record how many people voted for each division.}
\item[Printer:] retrieves and prints appropriate ballot, including SerialNumber and district, with private WBB signature.\footnote{Details on what is signed and how are in Section~\ref{sec:PoD}}
\item[Voter (Check 1): ] may choose to check and confirm this ballot.  This involves demanding a proof that the ballot is properly formed, {\it i.e. } that the permuted candidate list corresponds correctly to the ciphertexts on the public WBB for that serial number.  If the ballot has a 
confirmation check, the voter returns to the printing step for a new ballot.
\item[Voter: ] shows the printed ballot barcode to the EBM, then enters the vote via user-friendly EBM interface,
\item[EBM: ] prints on a separate sheet: 
		\begin{enumerate}
		\item the electoral district,
		\item the SerialNumber,
		\item the voter preferences permuted appropriately to match \pret{} ballot,
		\item a QR code with this data, plus private WBB signature.
		\end{enumerate}  
		
		\emph{This is the voter's receipt.}  An example is shown on the right side of Figure~\ref{fig:ballot}.

\item[Voter (Check 2): ] checks printed preferences against printed candidate list, and checks that the district is correct and the SerialNumber matches that on the ballot form.  
\item[Voter (Check 3):] (optionally) checks the WBB signature, which covers only data visible to the voter.  This requires an electronic device.
\item If either of Check 2 or 3 fails, the vote is cancelled using the cancellation protocol of Section~\ref{subsec:cancel}.
\item[Voter: ] shreds the candidate list,
\item[Voter: ] leaves the polling place.
\item[Voter (Check 4):] later checks their vote on the public WBB.  They only need to check the serial number and order of their preference numbers.
\item[Anyone (Check 5): ] after polling closes, checks the universally verifiable proof that all submitted votes are properly shuffled and decrypted.
\end{description} 

The rest of this Technical Report expands on each of these steps so as to give a complete account of end to end verification and an analysis of privacy. 

\section{Security Properties}\label{sec:security properties}
The intention is to provide a proof of integrity independent of any trusted hardware, software or people, while preserving reasonable privacy.

We have several different kinds of security assumptions, which apply at different points:

\begin{description}
\item[Computational assumption: ] A reduction to a computational problem generally believed to be hard.  For example, the privacy of ElGamal encryption relies on the hardness of the Decision Diffie Helmann problem on the elliptic curve being used.  The soundness of the zero knowledge proofs of correct decryption, which use the Fiat-Shamir heuristic, relies on the assumption that inverting the hash function is as hard as inverting a random oracle.\VTNote{Not sure this is well phrased.  SAS: seems fine to me.}
\item[Auditing assumption: ] An assumption that a sufficiently large and unpredictable fraction of a set have been confirmed or checked.  For example, proper ballot generation, proper ballot printing, and accurate printing of voter preferences all need to be checked with high enough probability and unpredictability to give us confidence in the accuracy of those that were not checked.
\item[Threshold or distributed assumption: ]  An assumption that a known threshold of authorities will not misbehave.  For example, votes on the public WBB are private as long as fewer than a threshold of the authorities who share the decryption key collude, and not all the mixers collude.  Robustness and reliability of the private WBB are also dependent on threshold assumptions.   The design requires a threshold greater than $2/3$ of the number of peers.  For example, if there are $7$ peers then a suitable threshold is $5$.

\item[Individual trust assumption: ] Trusting a single device.  For example, the EBM a person uses to vote is trusted not to leak the vote.  The printer is trusted not to leak ballot information.
\end{description}

Obviously, the intention is to minimise instances of trusting a single device.  Our system design aims to provide
\begin{description}
\item[Integrity ] based only on computational or auditing assumptions.  If a sufficient number of confirmation checks are properly conducted then the election's integrity is demonstrated given only computational assumptions.\footnote{The computational assumptions are due only to the use of the Fiat-Shamir heuristic to choose challenges for the proof of a shuffle.  An alternative method of generating unpredictable challenges based on some other assumption could also be used, in which case there would be no computational assumptions for integrity.}
\item[Non-repudiation ] based on a threshold assumption.  Unless more than a threshold of private WBB peers collude, it should be infeasible to produce a properly signed receipt without properly casting a vote.
\item[Robustness ] based on a threshold assumption.  If a threshold of private WBB peers behave properly, a properly signed receipt is guaranteed to appear on the public WBB.
\item[Privacy ] is the most subtle property, and needs to be discussed separately at several points.  We assume that the link between an individual and their receipt is public (though names are not printed on the WBB).
	\begin{description}
	\item[The printer ] is trusted not to leak ballot information via side channels. 
	\item[The EBM ] is trusted not to leak the vote via side channels.
	\item[The printer ] is prevented from performing kleptographic attacks by the ballot generation confirmation check.  The proper generation of randomness for those ballots depends on at least one of the randomness generation servers being honest.
	\item[The encrypted votes on the WBB ] remain private under threshold assumptions on the decryption key sharers and an assumption that there is at least one honest mix server.\VTNote{More precise.  SAS done}
	\item The system does not as it stands defend against pattern-based coercion attacks (a.k.a. ``Italian attacks''), or other subtle coercion techniques such as forced randomisation.
	\end{description}    
\end{description}

Verification of voter eligibility is dependent on human procedures: the system is dependent on a secure procedure for ensuring that only eligible voters can vote, with at most one vote each.  We assume that at some point a ledger of how many people have voted in each division at each polling place is reconciled with the published list of encrypted votes on the WBB.

We emphasise that for verification purposes all the voters and other observers have access to the public WBB, which is broadcast on a reliable channel.  Hence there are no  human or electronic components which must be trusted for integrity (apart from voter eligibility and polling-place ballot stuffing).  \VTNote{Update}

There are, however, threshold trust assumptions for liveness, reliability, and non-repudiation.   In other words, we rely on certain thresholds to prevent certain kinds of failures, although all those failures would be detectable even if all the authorites misbehaved.  (Whether they would in practice be detected might depend on an auditing assumption.) The private WBB peers provide a robust database implementation that distributes trust so the electoral authorities will only publish something that is verifiable as long as the trust assumption holds.  Since the authorities are responsible for choosing the peers to trust, they are responsible for meeting that assumption.  

Another way of looking at it is that the voters themselves do not need to trust any person or software for integrity, because they can verify it.  The authorities want to be confident that what they publish will indeed verify.  The design tells the authorities that, under certain trust assumptions, the system will give them what they need, and hence satisfy the interested and sceptical members of the public who want to verify the outcome.

\SASNote{Language changed slightly - rather than talk about the public not trusting the system, I prefer to talk about members of the public who want to verify.  Analogy with banking: I trust my bank, but still look at my statement}

The confirmation checks involved in verifiability also provide a way of catching bugs or errors in the software:  a failed check might also be due to a coding error, and successful checks also demonstrate the absence of coding errors that could affect the result of the election.

The protocol uses digital signatures to provide evidence of many kinds of failures (in addition to the more traditional end-to-end verifiability literature, which tends to focus on detection alone).  This provides two kinds of benefits: voters can prove that a malfunction occurred, but not persuade anyone that a malfunction occurred when it did not.  This is important in defending against the ``defaming attack'' in which people pretend to have detected a system failure which did not actually happen. Of course, there can be no proof that the EBM accurately represented the voter's intention: that step is dependent on the voter's testimony and hence is to an extent vulnerable to the ``defaming'' attack.

\VTNote{Add comment from CB:
I put a lot of emphasis on bug detection via the confirmations as well as the deterrent value of them.  I have not got a lot of bites
saying that certain attacks can occur since these are either not understood or assumed to be very highly unlikely.  So I call it
 "DDD" Detect, deter and defend.  Verification is heavy on the first two Ds.  I think James would like this.  SAS: added the paragraph before last about coding errors.
}

\chapter{Procedural details}\label{sec:polling-place-procedures}
This section details, from the human perspective, how certain important security conditions are enforced by insisting on particular human procedures.  (The next chapters explain how electronic processes guarantee other security properties.)  The most important procedures in the polling place include giving each voter exactly one, correct, ballot, allowing them to chose some at random to perform a confirmation check on, encouraging them to check their printed vote and its signature, and insisting that they shred their candidate list.  The procedures and guarantees for vision impaired voters are slightly different from those for sighted voters, because checking the printout requires the use of a device.

This chapter describes what checks should be performed to test for normal operation.  Recovery from failures is described in Chapter~\ref{sec:robustness-recovery}.

\section{Typical voters}

\subsection{Getting a ballot}
The voter presents herself to an official at a polling station and her name is 
marked off a register. The official sends the print station a request for a 
ballot of the appropriate LA and LC division. The print station prints the ballot 
with a Private WBB signature.  (Print station is abbreviated as VPS in other documentation.)

\emphbox{It is essential for privacy that no-one 
except the voter sees the association between the candidate order and serial 
number on the ballot, so printing should be private.}

Obviously it is essential for integrity that each voter is allowed to vote at most
once, on a ballot of the appropriate division.  This must be enforced by 
procedures at the polling place.

\emphbox{Reconciling the number of marked-off voters in each division with the number posted on the WBB is essential for preventing ballot stuffing.}

{\bf Check 1a: Confirming ballot correctness.}  
Once she has obtained her ballot, the voter should decide whether she wishes to run a 
confirmation check on it or use it to vote. A confirmation check, called 
``auditing'' in previous versions of \pret{}, means checking that the encrypted 
list of candidates on the WBB matches the plaintext candidate ordering on the 
RHS of the ballot. Ballot
confirmation ensures that the ballot is well-formed and hence
 would correctly encode a vote. We describe the ballot confirmation procedure 
below in Section~\ref{sec:const}. She can repeat the ballot confirmation 
procedure as many times as she wants in principle, each time obtaining a fresh ballot, until 
proceeding to vote using the last obtained, unconfirmed ballot. This implements 
an iterated cut-and-choose protocol: not 
knowing which option the voter will choose before committing to the printed 
ballot serves to counter any attempts by the system to manipulate votes by 
issuing malformed ballots. Confirming ballot construction necessarily reveals 
encryption information, so a ballot that has been confirmed should not be 
subsequently re-used for voting. 

\emphbox{It is essential for integrity that all voters have the opportunity to
perform a confirmation check on as many ballots as they wish.}  The more
voters who check, the stronger the evidence that the ballots are well formed.

{\bf Check 1b: Checking the WBB signature on the printed ballot.}
Each ballot is printed with a WBB signature to indicate it is legitimate.  Voters 
should check this signature before voting---if they vote on an illegitimate
ballot {\it i.e.} one which did not originate properly from the PoD protocol, their
vote will not be accepted and their privacy could be breached.  This does not
in itself affect integrity, because an attempt to deprive a person of a vote by
giving them an illegitimate ballot will be detected at voting time when the
WBB refuses to sign their submitted vote.

\subsection{Casting a vote} \label{sec:casting}
Assuming that she is now happy to proceed to casting her vote, the voter takes 
the last obtained ballot to the booth. In standard \pret{} she would now proceed 
to fill in her preferences directly on the ballot. However, given that the 
LC-BTL section contains about 30 candidates, it is not reasonable to expect the 
voter to enter her ranking preferences using a permuted candidate list. Instead 
we propose to use a touch screen Electronic Ballot Marker (EBM) that will 
display the candidates in standard order, as previously introduced in 
\cite{EVOTE2012:VEC}.  The voter enters her preferences via the screen in the 
standard way, then the EBM permutes them to match the candidate permutation
on her ballot. 
This means that we have to sacrifice one of the pleasing features of standard 
\pret{}: that no device directly learns the voter's choices. This seems 
unavoidable for such expressive ballots if the system is to be usable.

She inserts the ballot into the EBM and selects her preferred language and can 
run through a training module on the machine to learn about the whole voting 
procedure, verification and tallying. The voter is now offered the choice of 
sequence in which she votes that is, the Legislative Assembly (LA) or 
Legislative Council ballots, and for the latter she can vote either ``above the 
line'' (ATL) or ``below the line'' (BTL).  Note that although the voter 
can vote at any polling station, the LA ballot is specific to the region in 
which she is registered. She must however, fill in both a LA and LC ballot 
and will be prompted by the EBM to ensure that she does this.\footnote{Exact
rules on ballot spoiling are a matter of user
interface: at present, voters are allowed to cast incomplete or invalid
preference lists, as long as they are warned.  The receipts then reveal
their decision to spoil their ballot. An alternative, but not currently implemented, method, would be to include a candidate
called ``spoiled ballot'' who would be the first preference of any invalid
ballot.  Subsequent preferences would be meaningless, but could be filled in to
make the receipt look like that of a valid vote.  This would hide whether
the voter had voted formally or not.}

For each ballot (LA or LC), the EBM scans the the QR code which represents the 
permutation of the candidate ordering on her ballot and displays the candidates 
in legal ballot order. Once the voter enters her choices, she is asked to confirm 
her choices and when she does so, the EBM prints on a separate sheet of paper:
		\begin{enumerate}
		\item the district,
		\item the SerialNumber,
		\item the voter preferences permuted appropriately to match the \pret{} ballot,
		\item a QR code with this data, plus private WBB signature.
		\end{enumerate}  
This is the voter's receipt.
Note that the EBM knows the permutation on the ballot and so re-orders the 
voter's selection accordingly.  Note also that the EBM can assist the 
voter by pointing out syntactic errors, for example, duplicate rankings etc.

Before printing, the EBM submits to the Private WBB exactly the data it will 
print on the receipt. Then at printing time it adds the Private WBB signature, as 
a further QR code, onto the receipt.

{\bf Check 2: EBM vote printing.} The voter should check that the printed receipt
matches her intended vote.  This includes checking that the serial numbers match, 
and that the printed preferences match her intended vote arranged according to the 
candidate order on her ballot.

\emphbox{It is essential for integrity that all voters are encouraged to check that
their printed vote matches their intention.}

The voter now \emph{folds her candidate list} to keep it secret, and leaves the booth with both pieces of paper.  There should be a public space inside the polling place that allows officials to enforce the following procedures without exposing voters to coercion.

Existing laws preventing 
voters from photographing their ordinary paper ballots should also apply to 
the candidate list, for the same reason: a voter who retains evidence of the
order the candidates are listed on her ballot can prove later how she voted.

{\bf Check 3: Private WBB Signature on vote.} The voter can check the signature using a purpose-built smart phone app.  This
 must of course incorporate a check that the data signed by the WBB is the same
 as the data printed on the paper.

\emphbox{It is essential for non-repudiation that the voter checks the signature
on her receipt before leaving the polling place.}  If she fails to check, and does 
not receive a properly-signed receipt, then she will be able to detect later, but not
to prove, that her properly-submitted vote has been excluded from the WBB.

This is the voter's last opportunity to cancel her vote, for example if Check~2 or Check~3 have failed, or no receipt has been issued.  Procedures for vote cancellation are described in Sec~\ref{subsec:cancel}.

Next, the voter shreds the candidate list.  
This prevents her from proving how she voted.

\emphbox{It is essential for privacy and integrity that all voters are required to shred their
candidate list before leaving the polling place.} 

The voter should be easily able to produce multiple copies of her receipt, for example using
a photocopier or a camera (on a smartphone).  This
combats the ``trash attack,'' \cite{Benaloh:trashAttack} and also allows others to check her receipt on the
WBB.  It would also be quite reasonable for the VEC to retain duplicate copies of 
receipts, as well as letting the voters take them home.  Of course there would have
to be a careful procedure for ensuring that the centrally retained reciepts were 
accurate copies of the voters'.

{\bf Check 4: Receipt appears on WBB.}
After a given time period,  the voter can use her receipt to check that the 
information is correctly recorded on the WBB.

These 4 checks provide evidence that the vote is cast as the voter intended, 
and included unaltered in the count.

We now describe the ballot confirmation process in more detail.

{\bf Check 5: Checking the mixing and decryption proofs on the WBB.}
Anyone can verify the single, public, proof that all votes are correctly mixed and decrypted.

These 5 checks provide evidence that the vote is cast as the voter intended, 
and included unaltered in the count.
We now describe the ballot confirmation process in more detail.

\subsection{Confirming ballot correctness} \label{sec:const}
{\bf Check~1a: confirming ballot printing}
To perform Check~1, confirming ballot correctness, the ballot can be taken back 
to the printer. The printer prints a proof of correct ballot formation, along with a WBB signature.  The WBB must record that the ballot has been confirmed, and therefore not accept 
any vote cast with that ballot form. As part of the confirmation process, a 
clear ``CHECKED---NOT TO BE USED TO VOTE'' message (which must be visible) is printed on the ballot form.

The voter can also check the proof of decryption later on any other machine,
including at home, so we are not trusting the polling-place machines for
confirmation of ballot construction.
 
When the day's WBB becomes available (see Section~\ref{sec:wbb}), it 
shows which serial numbers were confirmed and displays a proof of what the
candidate ordering should be.  (It also shows which ones were voted and what the
preferences were.)  

Ensuring the mutual exclusion of confirmed and cast ballots is vitally important.
The Private WBB must run a realtime check that the same ballot is not both
confirmed and 
voted.  This process is
trusted for privacy, but not for integrity because violations are detectable. 

{\bf Check 1b: Verifying the WBB signature on the printed ballot.}
Each ballot is printed with a WBB signature that includes its Serial Number and district, to indicate it is legitimate and has been registered for the correct district.  Voters should check this as part of ballot confirmation.   (This prevents a corrupt printer from printing candidates for one district onto a ballot paper that is actually registered for a different, presumably more marginal, district.)

Check~1b could be performed on any ballot, including those that will be used for voting on.  However, it is difficult to allow this while also enforcing procedures for preventing voters from recording their candidate list.  In the absence of such procedures, Check~1b is only part of the ballot confirmation procedure.

In Section~\ref{sec:robustness-recovery}  we describe what to do when some
of these checks fail.

\subsection{Ballot cancellation: individual quarantine} \label{subsec:cancel}

There may be legitimate circumstances when a voter finds a check is not successful and wishes to instruct that the vote should not be cast. This may be used for various failures, including
\begin{itemize}
\item when a voter claims the printed vote differs from their intention
\item when a printed vote does not include a valid WBB signature
\item when the EBM fails to produce a printout
\item when an attempted ballot generation  or ballot printing confirmation check fails, either because it times out or because it does not produce a valid proof of correctness or a valid signature.
\end{itemize}  

The terminology used by VEC for this process was {\em Individual Quarantine (IQ)}.

A cancellation request overrides any other request, such as confirming or voting.  When a vote is cancelled, the cancellation is recorded against the Serial number on the (private and public) WBB.  The voter must give up their candidate list  in order to request a cancellation.

\emphbox{A cancellation request is allowed only if the voter presents the candidate list, and never after the voter has left
the polling place.}

It is important to emphasise that a vote is never cancelled except according to the following procedure. 
The process is:
\begin{enumerate}
\item	The voter requests a cancellation and provides the candidate list.  If the candidate list is already shredded or missing, then cancellation is refused.
\item	Polling official scans the Serial Number on the ballot and requests a cancellation.  
\item	VEC HQ provides permission for the cancellation to occur.  This authorisation of cancellation is uploaded to the WBB, which replies to the printer with a receipt.  Printer prints a signed cancellation onto the ballot.
\item Voter checks signature on cancellation.
\item Polling officials make a paper log of the cancellation, which is signed by the voter and retained by the electoral commission.\footnote{We would like to be able to guarantee that people cannot walk out of the polling place with validly signed receipts that have nevertheless been cancelled; unfortunately, this cannot be enforced---voters can always pocket their valid receipt and claim they never got one.  We need to be careful that they cannot cancel it and then use their preference printout to claim that their vote was incorrectly cancelled.  The insistence that they sign a paper log of their cancellation request is designed to defeat this attack.}
\end{enumerate}

The intention and expectation is that this process is used rarely, and with the explicit observation by at least two (and preferably more) officials at the polling place.  Cancellation requests should be independently recorded on paper at the polling place, and should require approval from senior officials.  If the voter does not have a ballot receipt, for example because the EBM failed to produce a printout, then this will also be recorded.

\VTNote{Still hazy on this one, but since its requirements seem to change by the day I'm relaxed about leaving it out for now.}

\VTNote{I'm not going to call cancellation ``quarantine,'' which has
an entirely different connotation.  Quarantine is for things that are suspect but, by default, are allowed through after a period of checking that they do not develop any further worrying symptoms.  These votes are cancelled.  They're not going to be counted, and many of the situations in which they're counted are precisely those in which we're not sure whether we've correctly received them anyway.}
\SASNote{I discussed this with Chris, and there's now a protocol for cancellation, and a rogue Cancel Station with the right serial numbers can post cancellation requests.  If it does too many then it may be identified as malicious, and then the cancel requests might be revoked.  All this would be on the record, so the cancel requests would be on the WBB, and then an explanation of where the judgement came from to ignore them.  So the word `quarantine' is less ridiculous in this context.  I have tried to write this section without using the word 'cancellation' too much just because it makes the VEC so jumpy, but it reads oddly at the moment.}

\subsection{Procedures for defeating chain voting} \label{subsec:chain}
Chain voting, which applies to conventional voting too, is an attack in which a coercer smuggles a (partially) completed ballot out of a polling place, and then
gives it to a voter with instructions to cast it and bring an unmarked
ballot back out.  The coercer then fills in this ballot and uses it to repeat the chain voting attack with a new voter.  

In \pret{}, the chain
voting risk applies to a coercer who obtains a printed ballot form, 
records the candidate order, and then sends a voter into back into the 
polling place with instructions to vote in a particular way and return
with both a receipt and a new, unmarked, ballot form.  Since the coercer
has already recorded the voter's candidate order, the receipt shows
how the person voted.  The new ballot form is used to repeat the attack with a new voter.

vVote includes some technical measures to defend against chain voting.
Printed ballot forms expire after 5 minutes if they have not been used to 
start a session, and the private WBB refuses to allow the same ballot form to be used to start another voting session once it has been used to start one.
This means someone who sneaks an unused printed ballot form out of the 
polling place has 5 minutes to send it in with another voter.  If someone sneaks one out having used it to start a session (and the tablet sits there with session active), then attempting to sneak this back in will not work as the printed ballot cannot be used to start a fresh session and the abandoned session itself ``locks.''.

\section{Vision impaired voters}
\label{subsec:blindCasting}
We assume that the vision-impaired voter has registered at a polling place and
had her name marked off.
The printing station should work fine for vision impaired voters, though there may need to be special procedures to help them collect their ballot privately.  If they need assistance, it is important that the assistant does not see the printed candidate list.  

The vision impaired voter takes the slip to an Electronic Ballot 
Marker (EBM).  
At the EBM, she inserts the slip. The system is set so 
that she has an audio-only session in her preferred language and the touch 
screen is laid out like a keypad, following the TVS2 standard. For example, the four corners when touched 
render 1, 3, $\ast$ and \#, the middle top and bottom give 2 and 0, and so on.\footnote{See 
{\tt http://www.eca.gov.au/research/files/telephone-voting-standard-reviewed.pdf} We have also made a new kind of interface, similar to (but invented before) this:
{\tt http://appleinsider.com/articles/13/03/26/\\\phantom{http://ap}apple-patents-no-look-multitouch-user-interface-for-portable-devices}}
 
The session is similar to the one described previously in 
Section~\ref{sec:casting} in that the voter has to fill in ballots for her LA 
and LC (ATL/BTL) votes, but this time she indicates her choices by touching the 
appropriate parts of the screen and has voice prompts to guide her. When she 
has filled in all required parts of the slip, she is given a voice confirmation 
of her vote choices and if she agrees with them, she can finish the voting part 
of the ceremony by touching the designated part of the screen.

As before, the EBM prints her reciept, including the SerialNumber, division, 
preferences, and Private WBB signature. 

This voter is unable to perform by sight the crucial check that the printed 
values match her intended vote.  Hence she may take both her candidate list and printed preferences to another EBM, which scans the QR code and the printed preferences, and reads her vote back to her.   This service is called ``The Readback App.'' It can also read back just a preferences list or just a candidate list.

\emphbox{It is essential for integrity that vision-impaired voters are encouraged
to check their printed preferences using an independent machine.}  If a compromised EBM can predict that a particular voter will not check
their preferences, the vote can be manipulated.  This 
cast-as-intended verification mechanism is requires trusting that the voter can find at least one EBM in the polling place that does not collude with the first one she used.

An alternative design would be to allow voters to bring their own devices in to
perform this check, but this would violate vote privacy because the device 
might record the data, hence telling someone else how the person voted.

By this point we can be confident that the printed preferences match the voter's
intentions.  She must now destroy the candidate list.

\emphbox{It is essential for privacy that all voters are required to shred their
candidate list before leaving the polling place.}

As already mentioned, the EBMs can also speak the preference orders on the slip so the voter can 
note them down (with a blind note-taker device or with memory). This helps the 
voter to check the EBM unassisted but does not really affect 
privacy or verifiability because she must still check that her vote is printed as she requested, and recorded
on the WBB as it is printed, rather than trusting the EBM to tell
her the truth.  She could do the WBB check with assistance from a print reader
or from a sighted person without jeopardising privacy. 

Note that the only steps that need to be private are the ballot marking by 
the EBM and check with a second EBM. All the other verification steps: 
confirmation of the ballot, confirmation of the receipt signature and of 
correct posting of the receipt to the public WBB, are exactly the same as those
for typical voters, and can be performed with assistance 
without jeopardising ballot privacy.

\commentOut{
Knowing
their first preference is 4 candidates down the list is necessary to
checking what ScanStation 
says. Maybe the voted-on EVM can casually point out that "the order of
your preferences on the voting receipt will be different to the actual
ballot. On the Northcote ballot, your first preference choice Sandy
Stone is seventh on the list. On your receipt Sandy Stone is forth.
However, your first preference is still shown for Sandy Stone and so are
the other preferences you cast. This shuffle of the candidate order is a
measure to protect the secrecy of your vote and allow you to take home a
receipt with your actual vote on it."

3.      At ScanStation <https://wiki.cse.unsw.edu.au/vec/ScanStation> ,
it will show the scan and the OCR result. Potentially the OCR result
could be read out. The voter would need to know her first preference in
the LA was four candidates down the list. The OCR result will only be
"Legislative assembly voting receipt. 5, 3, 7, 1, 4, 2, 6...". Voter
knows it is legit because her 1st pref is in position 4.
}

\subsubsection{Confirmation}
If she has
performed a confirmation check on a ballot, the voter can still go home and
use her screen- or print-reader, with the same confirmation-checking software as
everyone else, to make sure her candidate list matches the encrypted
list on the WBB.  The only
important detail is that she has to make sure she knows what the cleartext
candidate order is.  She must either ask several people or use (a) print
reader(s).  Neither of these impacts upon privacy: there are no privacy implications 
for anyone in confirming ballots.

\section{Observing that the vVote output matches what is 
input into the count} \label{sec:vVoteOutputIntoCount}
VEC procedures require vVote ballots to be printed out before being
incorporated into either the manual tally of paper votes (for LA ballots)
or the manual data entry of paper votes into the electronic STV count
(for LC ballots).  The scrutineers who observe the manual tally must
reconcile these ballots with those output from vVote on the public WBB.  

The printouts will be visually distinct from ordinary paper ballots.
All the vVote votes will bear a unique number on their footer which aligns with a verifiable output vote on the WBB so 
that they can be checked independently later.  Note that these unique numbers are added to the votes \emph{after} they have been shuffled and hence disassociated from the voter who cast them.

\SASNote{I understand that all votes are typed into the system and published (providing anonymity sets are large enough).  If vVote votes are published with their unique numbers then they could be checked by anyone against the output from the decryption.  This is just checking that information has been copied across correctly.  But I don't know if votes entered into the system can be identified as having been voted electronically, and whether the unique numbers can remain with them.  Need to check with Craig} 

\chapter{System Component details}\label{sec:system-component-details}
\section{The Web Bulletin Board}\label{sec:wbb}
A number of voting schemes require some form of append-only Web Bulletin Board
(WBB).  However, specific details of how to design or implement such a service are
often lacking. In this section we do not aim to propose a generic WBB, only to
define one that will work within the constraints we have and offer the
properties we need. The fundamental
requirements we have of a WBB are
\begin{itemize}
  \item that every observer gets the same information, and
  \item that the data written to it cannot be changed
or deleted without detection. 
\end{itemize} 

In prior work on the \PaV{} protocol, a great deal has been expected of the WBB. 
It has been expected to prove that it satisfies the properties above, while also
responding in real time with signatures and confirmation check information.  The design
presented here separates those two sorts of roles, breaking the WBB into two:

\begin{description}
\item {\bf The Public WBB } This is a static digest of the day's transcript.  It is
updated very infrequently (e.g. once per day).  A hash is broadcast via some
other channel (e.g. a newspaper or radio broadcast) so everyone can be confident
they all get the same information.  This data could be replicated extensively
because no secret information is held.  This is what voters consult for evidence
that their votes were included in the tally, and evidence that the tally was
correct.  The main property is: \emph{A corrupt WBB cannot falsify its data
undetectably except by violating some computational assumption.}  
\item {\bf The Private WBB } A robust secure database which receives messages, performs basic validity checks, and returns a signature.  Its correctness based on threshold assumptions.  This is implemented as a collection of peers who share a signing key.  Validly signed messages are guaranteed, under threshold assumptions, to appear subsequently on the Public WBB.  A malicious threshold can collude to misrecord and expose votes.  Such misbehaviour is detectable in principle by observing the public WBB, but may not necessarily be provable.  
\end{description}

The result of this split should be that the Private WBB can follow a protocol for distributed secure databases without needing to worry about reliable broadcast to the public.  The real broadcast channel with memory is a static data structure,
the WBB, which is much easier to design.
In practice the WBB could be replicated in the cloud. The
Private WBB
would receive communication only from other ``inside'' entities such as print
servers and EBMs.  

\subsection{Public Web Bulletin Board (Public WBB)}
The Public WBB is an authenticated public broadcast channel with memory.  We assume some genuine public broadcast channel that can be used to send a small amount of information, specifically a signed cryptographic hash of the transcript.  When someone checks the public WBB for inclusion of their data, they also re-hash the contents and check the result against the publicly broadcast one. 

The public WBB consists of static data broken into separate commits, with the signed hash of the prior commit step being included in each commit along with the other election data.

Verifying that something appears on the public WBB is a two step process:
\begin{itemize}
\item Firstly the observer requests an index file that lists all ballot serial numbers (or whatever other data is being requested) and which commit they are in. The client side code allows the voter to look up the serial number and then download the relevant commit. 
\item The second step is downloading all the data for the relevant commit, checking that a particular data item (e.g. a particular voting receipt) is present, and checking that the recomputed hash matches the signed, published one.
\end{itemize}

If the total amount of data per commit becomes large in future, the process of proving inclusion could be made more efficient using hash trees or other log-size data structures \cite{DBLP:journals/algorithmica/GoodrichTT11}.

\VTNote{btw, I do still like Craig's idea of presenting a separate pretty
(human-readable?) version, and I'd suggest that the verifier be designed to take the pretty version and compare it against the proofs.}

\subsection{Private Web Bulletin Board (Private WBB)}

A private WBB is a robust distributed database which:
\begin{itemize}
\item accepts items to be posted (if they do not clash with previous posts), 
\item issues receipts (which are signed accepted items), and 
\item periodically publishes what it has received on the public WBB.  
\end{itemize}

The data published on the public WBB for any particular period must include all items that had receipts issued during that period.   Robustness is achieved through the use of several peered servers which cooperate on accepting items, issuing receipts, and publishing the bulletin board.  They make use of a threshold signature scheme which allows a subset of the peers at or above a particular threshold to jointly generate signatures on data.   The peers collectively provide the bulletin board service as long as a threshold of them are honest, and as long as a threshold of them are involved in handling any item posted to the bulletin board.  Thus the implementation is correct in the presence of communication failures, unavailability or failure of peers, and also dishonesty of peers.  The threshold $t$ required to achieve this must be greater than two-thirds of the total number $n$ of peers:  $t > 2n/3$.   There is no single point of failure: the system can tolerate failure or non-participation of any component, as long as a threshold of peers remain operational at any stage.  It also allows for different threshold sets of peers to be operational at different times.  For example, a peer may be rebooted during the protocol, thereby missing some item posts, and may then resume participation.  Details of the protocol are given in a separate publication \cite{SS14:WBB}.  

In addition to robustness and non-repudiation, the private WBB must also perform some basic validity checking on submitted items.  In particular, it must reject items that ``clash'' with previously accepted submissions: a new request to run a confirmation check or vote on a ballot of a given serial number must be rejected if that serial number has already been voted on.\VTNote{This needs to be more precise, I think.  Is the request to print separate from the request to vote/confirm?  Do we accept a print request from something that's already been voted/confirmed, or is that part of a ``clash'' too?   SAS response: Request to print is separate from posting or confirming of a vote.  I think that a second request to print would be rejected}

The key properties required of the Private WBB are:

\begin{description}
\item[(bb.1)] only items that have been posted to the bulletin board may appear on it;
\item[(bb.2)] any item that has a signed receipt issued must appear on the published bulletin board ({\it i.e.} public WBB);
\item[(bb.3)] two clashing items must not both appear on the bulletin board;
\item[(bb.4)] items cannot be removed from the bulletin board once they are published.
\end{description}
It follows from bb.2 and bb.3 that if two items clash then receipts must not be issued for both of them.

The bulletin board provides a protocol for the posting of an item by an EBM, and its acknowledgement with a receipt.  It also provides another two related protocols for the publishing of the bulleting board: an optimistic one, and a fallback.  These protocols are given in \cite{SS14:WBB} together with their proofs of correctness with respect to the key properties.

\section{Print-on-demand printers and Randomness Generation Service} 
\label{sec:PoD}
The processes for printing and confirming correctness of ballot forms are  vital components of \pret{}.  This project necessitated a completely new scheme, which is described below.

\subsection{Protocol overview}

Our protocol has two roles.  The ``randomness generation servers,'' of which a threshold of at least one are trusted for privacy, send randomness to a ``printer''.  The ``printer'' uses only that randomness to generate the ballots, which it can then print on demand.  In brief:

Before the voting period:
\begin{enumerate}
\item Each randomness generation server generates some randomness, commits to it publicly, and sends the opening secretly to the printer.
\item \label{step:ballotGen} The printer uses the combined randomness to generate the encrypted ballot, which it publishes.
\end{enumerate}

During the voting period:
\begin{enumerate}
\setcounter{enumi}{2}
\item \label{step:ballotPrint}  When required, the printer prints the next ballot in sequence, with human-readable candidate names.
\end{enumerate}

There are thus two important points for public confirmation checking:
\begin{enumerate}
\item A confirmation check of the encrypted ballot produced in 
step~\ref{step:ballotGen}, to check that the candidate ciphertexts are valid and that the printer used the proper randomness.  This is described in 
Section~\ref{subsubsec:ballotGenAudit}.
\item A standard \PaV{} confirmation check of the printed ballot from step~\ref{step:ballotPrint}, to check that the printed human-readable candidate names match those of the encrypted ballot.  This is described for our scheme in 
Section~\ref{subsubsec:printAudit}.
\end{enumerate}

Throughout this document when we refer to the ``Printer'' we are in fact referring to the tablet device that is connected to the printer. As such, the ``Printer'' has the processing power you would expect to find on a mid-range tablet.

\subsection{Ballot Generation} \label{sec:ballotGen}

The main idea is that the printer generates a permuted list of candidate ciphers using randomness values generated by a distributed set of peers. As such the printer undertakes the expensive crypto operations, but does not have any influence over the values used in those operations. This prevents the printer from mounting kleptographic attacks or otherwise having any influence over the ciphertexts.  

During ballot generation the printer is checked to ensure that it has performed honestly: if a sufficient number of ballots are confirmed and shown to be correct then we can gain a high assurance that the printer has behaved honestly.  The definition of ``honest'' is different from standard versions of \PaV, in which a dishonest printer can only misalign the printed candidate names with the ballot ciphertexts.  In our version, a dishonest printer may also attempt to generate invalid ballot ciphertexts or perform a kleptographic attack by using randomness other than that specified by the protocol.  However, these two kinds of cheating can be detected by a ballot-generation confirmation check---see below.

{\bf Notation: }
\begin{description}
\item[$\enc{k}{m}{r}$] is the encryption of message $m$ with public key $k$ and randomness $r$.\footnote{The {Victorian} project uses Elliptic Curve El Gamal.}
\item[$\dec{k}{m}$] is a decryption of $m$ using the private key $k$.
\item[$\renc{\theta}{r}$] is a re-encryption (re-randomisation) of the ciphertext $\theta$ using the randomness $r$ (this abstracts the requirement of knowing the public key, which is a requirement for ElGamal re-encryption).
\item[$\com{m}$] is a perfectly hiding commitment to message $m$ (using some randomness not explicitly given).
\item[$\com{m;r}$] is a perfectly hiding commitment to message $m$ (using randomness $r$).\footnote{In the {vVote} project we use the hash-based commitment scheme described in \cite{jakobsson02:e-vote}.}
\item[$\epk$] is the election public key (which is thresholded).
\item[$\ppk$]is the printer's public key (which is not thresholded/distributed).
\item[$\psk$] is the printer's private key.
\item[$n$] is the number of candidates.
\item[$\BC$] is the number of ballots to be generated for each printer.
\item[$\SC$]  is the number of randomness generation servers.
\item[$\Mix_i$] is the $i$-th randomness generation server.
\item[$\aes{sk}{m}$] is a symmetric-key encryption of $m$ under the symmetric key $sk$.
\item[$\SHA{m}$] is a cryptographic hash of the message $m$.\footnote{We use 256-bit AES and SHA-256 respectively.  This means that the computational difficulty of guessing a key ($2^{256}$) is much greater than that of finding a collision ($2^{128}$).  However, this seems justifiable since collision-finding is only useful for cheating during ballot generation, which must be performed in a restricted time, while guessing the symmetric key can be used to break ballot privacy long after the election.}
\item[$\aesdec{sk}{m}$] is a symmetric-key decryption of $m$ using key $sk$.
\item[$\\RT_i$] is $\Mix_i$'s private table of encrypted random values (Fig~\ref{tbl:BallotInput}).
\item[$\CRT_i$] is $\Mix_i$'s public table of commitments to the values in $\RT_i$ (Fig~\ref{tbl:committedBallotInput}).
\end{description}

We will post on the public WBB values that are encrypted with a threshold key, or perfectly hiding commitments. We will not post values that are encrypted with a non-thresholded key.  We could have used computationally hiding commitments or encryptions with non-thresholded keys, but either of these would have meant that a single leak of relevant parameters, even quite a long time in the future, could have been combined with WBB data to violate ballot privacy.  Our system does not achieve everlasting privacy, but it achieves a somewhat related weaker property, that no single entity's data (apart from the printer's) is enough to break ballot privacy, even given WBB data.

\begin{figure}
\centering
\begin{tabular}{|cc|}
\hline \bf Candidate Name & \bf ID \\
\hline
Vladimir Putin & $\cand_1$ \\ 
Mohamed Morsi & $\cand_2$ \\ 
 ... & ... \\ 
 Mahmoud Ahmadinejad & $\cand_n$ \\ 
\hline
\end{tabular} \caption{Initial Ballot Input: Candidate Identifiers} \label{tbl:CandIDs}
\vspace*{6ex}
\centering
\begin{tabular}{|c|c|c|c|c|}
\cline{1-1} \cline {3-3} \cline {5-5}
PrinterA:1 & \ldots & PrinterB:1 & \ldots & PrinterC:1  \\ 
PrinterA:2 & \ldots & PrinterB:2 & \ldots & PrinterC:2 \\ 
\vdots & \vdots & \vdots & \vdots & \vdots \\ 
PrinterA:$\BC$ & & & & \\ 
\cline{1-1} \cline {3-3} \cline {5-5}
\end{tabular} \caption{Initial Ballot input: Serial numbers for printers A,B,C.}
\label{tbl:InitialBallotInput}
\end{figure}

\subsubsection{Pre-Ballot Generation} \label{subsubsec:preBallotGen}
Before the ballot generation starts the following must occur:

\begin{enumerate}
\item The election public key sharers jointly run a distributed key generation protocol to generate a thresholded private key and joint public key $\epk$.\footnote{We keep the key sharers and the randomness generation servers conceptually separate, even if we end up using the same servers.}
\item A list of candidate identifiers is generated and posted on the public WBB, as shown in Figure~\ref{tbl:CandIDs}.  Candidate identifiers are arbitrary, distinct elements of the message space of the encryption function.\footnote{For {vVote}, candidate identifiers are elliptic curve points either randomly selected or calculated as part of the optimisation used to speed up mixing, depending on the type of race.}
\item For each printer, a list of serial numbers of the form ``PrinterID:index'' is deterministically generated and posted on the public WBB.  This serial number is just the literal string as given.  These serve as row indices for later computation. 
\item Each printer constructs a key pair and publishes the public key $\ppk$.
\end{enumerate}

All this data, which is posted on the WBB immediately before ballot generation, is shown in Figures~\ref{tbl:CandIDs} and~\ref{tbl:InitialBallotInput}.  The protocol for postsing a file to the WBB is shown in Figure~\ref{fig:FileMSC}.

\begin{figure}[th]
\begin{msc}{File Message Sequence Chart}\label{msc:file}
\setmscvalues{large}
\setmscscale{1}
\declinst{podt}{}{PODPrinter}
\declinst{wbb}{}{Private WBB}
\small

\nextlevel
\mess{boothID,boothSig,``file'',digest,fileSize,submissionID,desc}{podt}{wbb}
\nextlevel
\mess{submissionSig,submissionID,``file'',peerID,peerSig,commitTime}{wbb}{podt}
\nextlevel
\end{msc}

\begin{itemize}
  \item boothSig : {Sign$_{PODPrinter}$\{serialNo,digest\}}
  \item peerSig : {Sign$_{WBB}$\{submissionID,digest,senderID,commitTime\}}
\end{itemize}
	\caption{File Message Sequence Chart}
	\label{fig:FileMSC}
\end{figure}

The following sections describe the process of ballot generation for a single printer, but it should be clear how the same process will be run in parallel for each printer.

\subsubsection{Randomness Generation}
The randomness generation consists of each server $\Mix_i$ generating a large table of secret random values and sending them (privately) to the printer after posting (public) commitments to them on the WBB.   The protocol for posting the public commitments is illustrated in Figure~\ref{fig:MixRandComm}.

\begin{figure}[th]
\begin{msc}{Mix Random Commit}\label{msc:mixRandomCommit}
\setmscvalues{large}
\setmscscale{1.0}
\declinst{podt}{}{PODPrinter}
\declinst{wbb}{}{Private WBB}
\small

\nextlevel
\mess{boothID,boothSig,``mixrandomcommit'',digest,fileSize,submissionID,printerID}{podt}{wbb}
\nextlevel
\mess{submissionID,``mixrandomcommit'',peerID,peerSig,commitTime}{wbb}{podt}
\nextlevel
\end{msc}

\begin{itemize}
  \item boothSig : {Sign$_{PODPrinter}$\{submissionID,printerID,digest\}}
  \item peerSig : {Sign$_{WBB}$\{submissionID,senderID,printerID,digest,commitTime\}}
\end{itemize}
\caption{Mix Random Commit Message Sequence Chart}
\label{fig:MixRandComm}
\end{figure}

\paragraph{Detailed algorithm}
In detail: each random value has length $\kparam$, where $\kparam$ is a security parameter which should be about 256 bits.
Each server $\Mix_i$ generates a random symmetric key $sk_i$. It then generates a table of $\BC * n$ pairs of pieces of random data, each of size $2*\kparam$ bits and encrypted under $sk_i$.  The table is denoted by $\RT_i$, and each pair can be retrieved by the serial number and column, or the row and column.\footnote{For example $\Mix_i.\RT_i(PrinterA:1,2)$ and $\Mix_i.\RT_i(1,2)$ will return the pair of encrypted random values in the second column for the first ballot: $\aes{sk_i}{r_{1,2} || R_{1,2}}$.} The result is shown in Figure~\ref{tbl:BallotInput}. The idea is that the first element of each pair, $r_{(row,col)}$, will be used later by the printer; the second element, $R_{(row,col)}$, is used to commit to $r_{(row,col)}$ and to open the commitment in case of confirmations. 

\begin{figure}[t]
\centering
\begin{tabular}{|c|c|c|c|}
\hline SerialNumber & \multicolumn{3}{|c|}{Encrypted Randomness} \\ 
\hline
\hline PrinterA:1 & $\aes{sk_i}{r_{1,1} || R_{1,1} }$ &  \ldots & $\aes{sk_i}{r_{1,n} || R_{1,n}}$ \\ 
\hline PrinterA:2 & $\aes{sk_i}{r_{2,1} || R_{2,1}}$   & \ldots & $\aes{sk_i}{r_{2,n} || R_{2,n}}$ \\ 
\hline \vdots & \vdots & \vdots & \vdots \\ 
\hline PrinterA:$\BC$ & $\aes{sk_i}{r_{\BC,1} || R_{\BC,1}}$ & \ldots & $\aes{sk_i}{r_{\BC,n} || R_{\BC,n}}$ \\ 
\hline 
\end{tabular} 
\caption{Ballot Input: Table $\RT_i$, sent privately from peer $i$ to printer A without public posting.}
\label{tbl:BallotInput}
\end{figure}

Each server commits to $r_{(row,col)}$ by posting on the WBB a commitment to it using randomness $R_{(row,col)}$.  The table of commitments is shown in Figure~\ref{tbl:committedBallotInput}.  Call the table $\CRT_i$.   Each peer $\Mix_i$ posts its $\CRT_i$, and checks all $\CRT_i$'s are posted before sending $\RT_i$ privately to the printer.\footnote{This is to stop the last peer choosing their randomness when they know the others'.  If this was not enforced, then one bad randomness generation server colluding with a printer could determine the randomness values for each of that printer's ballots, thus breaking privacy.  The bad server would wait until the printer told it all the other servers' random values, then generate its own to produce a particular final value.}  $\Mix_i$ also   encrypts $sk_i$ with the printer's public key  and sends the result (denoted $esk_i  = \enc{\ppk}{sk_i}{r}$) to the printer.

\begin{figure}
\centering
\begin{tabular}{|c|c|c|c|c|}
\hline SerialNumber & \multicolumn{3}{|c|}{Committed Randomness} \\ 
\hline
\hline PrinterA:1 & $\com{r_{1,1};R_{1,1}}$  & \ldots & $\com{r_{1,n};R_{1,n}}$ \\ 
\hline PrinterA:2 & $\com{r_{2,1};R_{2,1}}$  & \ldots & $\com{r_{2,n};R_{2,n}}$ \\ 
\hline \vdots & \vdots & \vdots & \vdots \\ 
\hline PrinterA:$\BC$ & $\com{r_{\BC,1};R_{\BC,1}}$  & \ldots & $\com{r_{\BC,n};R_{\BC,n}}$ \\ 
\hline 
\end{tabular} 
\caption{Commitment to Initial Ballot Input: Table $\CRT_i$, posted by peer $i$ on the WBB.}
\label{tbl:committedBallotInput}
\end{figure}

\subsubsection{Ballot Permutation and Commitment}  \label{subsec:ballotGen}
The printer receives the respective $\RT_i$ and corresponding (encrypted) key $esk_i$ from each server.  The printer also downloads or constructs the candidate identifiers and serial numbers shown in Figures~\ref{tbl:CandIDs} and~\ref{tbl:InitialBallotInput}. The printer now needs to encrypt and permute the candidate identifiers.

For each candidate identifier $\cand_k$ in the pre-committed table in Figure~\ref{tbl:CandIDs} it encrypts it using the combined randomness from the $\RT$ tables, received from the randomness generation servers. To combine the randomness the printer first decrypts the encrypted symmetric key $esk_i$ received from each server and then uses the resulting key $sk_i$ to decrypt the randomness in $\RT_i$.  For each element of each $\RT$, the printer checks that the decrypted pair  $r_{row,col},  R_{row,col}$ opens the commitment at $\CRT_{row,col}$.  It challenges any that do not---see below for what it should do when it detects $\Mix_i$ cheating at this point.

The decrypted first elements $r_{row, col}$ from each peer are concatenated and hashed.
The printer then uses the hash output for randomness when encrypting the candidate identifier under $\epk$. The resulting ciphers are then sorted into canonical order to produce a random permutation. These ciphers are posted on the public WBB.  Note that the output of the encryption is pseudo-random and as such sorting the encrypted ciphers will give a pseudo-random permutation $\pi$. The printer retains this permutation so that it can print the plaintexts in the appropriate order when requested to print that ballot.  After the ciphers are submitted to the WBB,  the confirmation checking protocol detailed in Section \ref{subsubsec:ballotGenAudit} can be run.  The Ballot Generation Commitment protocol between the printer and the bulletin board is given in Figure~\ref{fig:BallotGenMSC}.

\begin{figure}[th]
\begin{msc}{Ballot Generation Commit}\label{msc:ballotGenCommit}
\setmscvalues{large}
\setmscscale{1.0}
\declinst{podt}{}{PODPrinter}
\declinst{wbb}{}{Private WBB}
\small

\nextlevel
\mess{boothID,boothSig,``ballotgencommit'',digest,fileSize,submissionID}{podt}{wbb}
\nextlevel
\mess{submissionID,``ballotgencommit'',peerID,peerSig,commitTime}{wbb}{podt}
\nextlevel
\end{msc}

\begin{itemize}
  \item boothSig : {Sign$_{PODPrinter}$\{submissionID,digest\}}
  \item peerSig : {Sign$_{WBB}$\{submissionID,digest,senderID,commitTime\}}
\end{itemize}
	\caption{Ballot Generation Commit Message Sequence Chart}
	\label{fig:BallotGenMSC}
\end{figure}

\begin{algorithm}
\center{\bf Algorithm 1: Deterministic Encryption by Printer}
\small
\begin{algorithmic}
\For{$i = 1 \to \SC$} \Comment{Decrypt the symmetric keys from the \Mix Servers}
\State $sk_{i} \gets \dec{\psk}{esk_i}$
\EndFor
\For{$j = 1 \to \BC$} 	\Comment{$\BC$ is the number of ballots.}
\For{$k = 1 \to n$} 	\Comment{$n$ is the number of candidates.}
\State $rand \gets SHA(\aesdec{sk_1}{\RT_{1}(j,k)}\|\ldots\|\aesdec{sk_{\SC}}{\RT_{\SC}(j,k)})$
\State $CT_{j,k} \gets \enc{\epk}{\cand_k}{rand}$
\EndFor
\State $CT_{(j)} \gets Sort(CT_{(j)})$  \Comment{$CT_{(j)}$ returns the entire row}
\State $perms_j \gets \pi$	\Comment{$\pi$ is the permutation applied to sort $CT_j$.}
\EndFor
\State Send $CT$ 
to WBB.
\end{algorithmic} \label{alg:detEnc}
\end{algorithm}

The algorithm run by the printer is given in Algorithm 1.
The intention is that only the printer knows which ciphertexts correspond to which candidates, but its algorithm for generating those ciphertexts is deterministic.  Hence it cannot use the ciphertexts to leak information without detection.  Of course, the printer could always leak that information via a side channel, but this is unavoidable and occurs with any form of electronic ballot printing or marking.

\paragraph{What the printer should do if $\Mix_i$ cheats} \label{subsubsec:mixCheats}
It is important in the above protocol that the printer checks the opening of each commitment, {\it i.e.} checks that for each element of each $\RT$, the decrypted pair  $r_{row,col},  R_{row,col}$ opens the commitment at $\CRT_{row,col}$.  It is important in practice that the printer raise an alarm on any commitments that are not correctly opened.  
Exactly how such a dispute should be resolved requires some careful engineering of procedures.  It is difficult to tell whether the printer or the randomness generation server is misbehaving without exposing private ballot data.  This is not necessarily a problem, because the randomness contributed by other randomness generation servers would not be exposed.  Hence the other ballots remain private.

Note that the issue does not affect public verifiability, because the absence of proper commitment opening would be detected by a confirmation check of this ballot. It does, however, affect accountability: if a confirmation check detects that the value used to encrypt a ballot was not a valid opening of the commitment on the WBB, we would like to know whether it was the randomness generation server or the printer that cheated.  If we insist that the printer performs this check, then we can be certain that a failed confirmation is the printer's fault.

\paragraph{Alternative construction with PRNGs}  \label{subsubsec:PRNG}
Rather than generate a separate random value for each candidate in each ballot, an alternative is to generate one random value for each ballot, then use a cryptographic Pseudo-Random Number Generator (PRNG) to expand it to produce randomness for all of the encrypted values in the ballot.
This introduces an assumption on the good expanding behaviour of the PRNG, but substantially reduces the communication costs of the protocol.  Hence the tables of Figures~\ref{tbl:BallotInput} and~\ref{tbl:committedBallotInput} would need only 2 columns rather than $n+1$.  It does not significantly change confirmation checking.

However, we have not taken this approach for the {Victorian} system due to practical concerns about PRNG implementations.  The most important is the possibility that variations in implementations could imply a failure of reproducibility of the random sequences, which is required for confirmation checking.  In principle this should not be a difficult issue to address.  In practice we were concerned this would make it significantly harder to write an independent verifier, so we opted to omit the PRNGs.

\subsubsection{Ballot Generation Audit}  \label{subsubsec:ballotGenAudit}
A suitable percentage of ballots are chosen at random for audit.\footnote{The questions of who chooses, how they choose, and how it can be guaranteed that they choose well enough to engender confidence in a particular election result are discussed in Section~\ref{subSec:auditCount}.} For each ballot selected, the printer posts on the WBB the randomness it used during the generation, {\it i.e.} to open the commitments for that SerialNumber in each peer's $\CRT$ table.  The protocol is shown in Figure~\ref{fig:BallotAuditCommit}. The printer can either have this stored, or else can recalculate it from the encryptions it received prior to ballot generation. Anyone can verify the commitment openings  $(r_{row,col}, R_{row,col})$ and reconstruct the ballot ciphertexts from them.  
Thus anyone can check that the ballots were correctly constructed and that the printer used the appropriate randomness. 

\begin{figure}[th]
\begin{msc}{Ballot Audit Commit}\label{msc:ballotAuditCommit}
\setmscvalues{large}
\setmscscale{1.0}
\declinst{podt}{}{PODPrinter}
\declinst{wbb}{}{Private WBB}
\small

\nextlevel
\mess{boothID,boothSig,``ballotauditcommit'',digest,fileSize,submissionID}{podt}{wbb}
\nextlevel
\mess{submissionID,``ballotauditcommit'',peerID,peerSig,commitTime}{wbb}{podt}
\nextlevel
\end{msc}

\begin{itemize}
  \item boothSig : {Sign$_{PODPrinter}$\{submissionID,digest\}}
  \item peerSig : {Sign$_{WBB}$\{submissionID,digest,senderID,commitTime\}}
\end{itemize}
\caption{Ballot Audit Commit Message Sequence Chart}
\label{fig:BallotAuditCommit}
\end{figure}

\subsection{Print on Demand} \label{sec:PoDPrinting}
This section describes what happens when a voter appears at a polling place.   The printer needs to print a pre-generated ballot for the appropriate district.  The printer knows the plaintexts and permutation for a particular ballot so can easily print the appropriate ballot out. 

\VTcomment{I'm strongly inclined to leave the rest of this subsection and the diagram out, or perhaps to give it a lot more background.  Not even sure whether the explanation I've added here is correct.  Why do we need a WBB sig anyway? It's hard to make sense of without the wider context of how the rest of the protocol works.  Anyway, it's in for now.  Comments welcome. }

However, there is a risk that a misbehaving printer might print a completely invalid ballot, {\it i.e.} one that has not been part of the generation process described above.  Although this is detectable by confirmation checks, we prefer for practical reasons to prevent it altogether.  Hence we require that the printer obtain a signature from the WBB in order to create an authentic ballot. The WBB is attesting to those ciphertexts matching what the printer has already committed to.  

The {vVote} project uses a combined EBM and scanner rather than the traditional \PaV\ technique of filling in the ballot with a pencil and then scanning it.  The print on demand protocol works exactly the same either way, but the description below assumes the {vVote}-style combined scanner and EBM.

\begin{figure}[th]
\begin{msc}{Print On Demand}\label{msc:pod}
\setmscvalues{large}
\setmscscale{1}
\declinst{vtr}{}{Voter}
\declinst{pwrk}{}{Poll Worker}
\declinst{podt}{}{POD Printer}
\declinst{wbb}{}{Private WBB}
\small
\nextlevel
\mess{\small ID}{vtr}{pwrk}
\nextlevel
\mess{\small district}{pwrk}{podt}
\nextlevel
\mess{\small boothID, boothSig, serialNo, ``pod'', district, ballotReductions}{podt}{wbb}
\nextlevel
\mess{\small serialNo, ``pod'', peerID, peerSig, commitTime}{wbb}{podt}
\nextlevel
\mess{\small peerSig, ballot}{podt}{vtr}
\nextlevel
\end{msc}

\begin{itemize}
  \item boothSig : {Sign$_{PODTablet}$\{serialNo,district\}}
  \item peerSig : {Sign$_{WBB}$\{serialNo,district\}} (the returned peerSig forms the serialSig in future sequence charts)
\end{itemize}
	\caption{Print on Demand Message Sequence Chart}
	\label{fig:podMSC}
\end{figure}

Figure \ref{fig:podMSC} shows the message sequence chart for print on demand, which is elucidated in Section~\ref{sec:balred} below.  The authorisation, ballot reduction and serial number signing all take place together in a single round of communication. 
The signatures generated are deterministic BLS signatures  \cite{boneh2001short},  including the signature from the WBB.

\subsubsection{Ballot Reduction} \label{sec:balred}
It is unpredictable exactly how many of each ballot will be required at each location.  We could have generated an abundant oversupply of ballots with exactly the right number of candidates for each division, but this would have been quite expensive.  Instead, for efficiency, we generate an abundant oversupply of generic ballots with a larger than necessary number of candidates, then reduce it down to the appropriately sized ballot for the district/region it is going to be used in.  This allows great flexibility about who votes at what polling place---any voter can arrive anywhere and have a ballot produced to match their voting eligibility.

If the ballot contains more ciphers than candidates, we need to reduce the ballot in a manner that can be verified. The following proposal has the nice feature that  the voter (or the EBM if there is one) does not need to know where the blanks are in order to cast the vote: they just get a permuted list of the candidates they were expecting.  Then the EBM  prints, and the voter checks, the voter's preference numbers against that order.  

Suppose from now on there are $m$ candidates in the division and $n$ $(>m)$ ciphertexts on the ballot. ($n=m$ is a special case of the steps below.) The printer is supposed to use the ciphertexts for $\cand_1, \cand_2, \ldots, \cand_m$.  Of course it could cheat and attempt to use other ciphertexts instead, but this could be detected at confirmation checking like any other kind of bad printing.  We want to be able to demonstrate afterwards on the WBB that it used the right ciphertexts.  The protocol is as follows:

\begin{description}
\item[Pollworker:] authenticates the voters (using whatever secure or insecure method is traditional) and sends a print request to the Print On Demand device specifying the district/region they can vote in,\footnote{More generally, it is the pollworkers' responsibility to authenticate the voter and request the ballot(s) that the person is eligible to vote on.}
\item[Printer:] retrieves the next available ballot and looks up the number  $m$ of candidates in the submitted district/region. 
\item[Printer:] sends to the WBB:
	\begin{itemize}
	\item the SerialNumber, 
	\item the division, and
	\item a list $BallotReductionRandomness$ of randomness values for the unused ciphertexts (i.e. the ones from $\cand_{m+1}$ to $\cand_{n}$), together with their respective permuted locations so the WBB can check them.  The randomness values are those computed by the printer in the algorithm in section~\ref{subsec:ballotGen})
	\end{itemize}
\item[WBB:] checks that the ciphers held for $\cand_{m+1}$ to $\cand_{n}$ are encryptions of the candidate IDs of the unused candidates for the specified division. 
\begin{itemize}
\item if valid it signs the serial number and the division and returns it to the printer, and posts the randomness values to the WBB so they can be publicly checked; 
\item if invalid it returns an error message.
\end{itemize}
\item[Printer:] Checks the WBB signature of the serial number and division and, if valid, prints the ballot and signature. The printer knows the permutation and plaintexts so does not need to do any crypto to print the ballot
\item[Voter:] votes on the ballot exactly as if it had been generated for the right number of candidates,
\item[EBM:] submits the ballot to the WBB exactly as if it had been generated for the right number of candidates,
\item[WBB:] accepts (and signs) the ballot only if it is accompanied by a signed serial number and division
\item[EBM:] prints the sig on the receipt,
\item[Voter:] (optionally) checks the sig, which covers only data visible  to the voter.
\item[Voter:] shreds the candidate list,
\item[Voter:] later checks their vote on the WBB.  They only need to check the serial number and order of their preference numbers---the correct opening of the unused (too big) candidate numbers will be universally verifiable.
\end{description} 

\subsubsection{Print confirmation}\label{subsubsec:printAudit}
Suppose a voter wants to confirm a printed ballot, i.e. to check that the printed candidate list matches the ciphertexts on the WBB. The following is performed:

\begin{description}
\item[Voter:] requests a confirmation check from the same printer that printed their ballot,
\item[Printer:]  sends to the WBB the randomness to open the commitments to the randomness on each $\CRT_i$ used to generate the ballot in the ballot generation phase
\item[WBB:] checks the serial number has not already been voted on or confirmed and if not, opens the commitments, reconstructs the ballot, computes the permutation $\pi$, posts all the data on the public WBB, and sends a jointly signed copy of $\pi$ (or candidate names in permuted order) to the printer
\item[Printer:] The printer prints the signature
\item[Voter:] checks the signed order of candidates $\pi$ against the order printed on the ballot (note, the permutation signed by the WBB should reflect any successful ballot reduction already performed).   
\item[Voter:] takes their confirmed ballot home and checks that the  value provided on the WBB matches the candidate order that was signed.
\end{description}

\begin{figure}[th]
\begin{msc}{Confirmation Check}\label{msc:audit}
\setmscvalues{large}
\setmscscale{1}
\declinst{vtr}{}{Voter}
\declinst{podt}{}{POD Printer}
\declinst{wbb}{}{Private WBB}
\small
\nextlevel
\mess{\small serialNo, district, serialSig, permutation}{vtr}{podt}
\nextlevel
\mess{\small boothID, boothSig, serialNo, serialSig, permutation, commitWitness, ``audit'', district}{podt}{wbb}
\nextlevel
\mess{\small serialNo, ``audit'', peerID, peerSig, commitTime}{wbb}{podt}
\nextlevel
\mess{\small serialNo, district, serialSig, permutation, peerID, peerSig, commitTime}{podt}{vtr}
\nextlevel
\end{msc}

\begin{itemize}
  \item boothSig : {Sign$_{PODTablet}$\{serialNo,``audit'',permutation,commitWitness\}}
  \item serialSig : {Sign$_{WBB}$\{serialNo,district\}}
  \item peerSig : {Sign$_{WBB}$\{``audit'',serialNo,reducedPermutation,commitTime\}}
\end{itemize}

	\caption{Print on Demand Confirmation Check Message Sequence Chart}
	\label{fig:podaud}
\end{figure}

Figure \ref{fig:podaud} shows the message sequence chart for confirming. 

\subsubsection{Forward Secrecy} \label{sec:forwardSecrecy}

The randomness held on a printer is sufficient to reconstruct the ballots, and hence reveal the candidate orderings.    If a printer is stolen, the randomness it holds will expose the associated ballot forms.  Hence it is desirable for a printer to delete the randomness for ballots on which votes have been cast, since there is a potential privacy breach.

However, when a printer prints a ballot it must allow for a print confirmation check which will require it to open the commitment to the randomness.  Therefore, after a printer has printed a ballot, it must retain this randomness for some period of time.

After the ballot has been used to cast a vote, a confirmation check is not allowed and so the randomness no longer needs to be retained and can be deleted.  Deletion can be triggered by a confirmation message to the printer, signed by the WBB, when a vote is cast.  Alternatively a time limit can be set on a confirmation request following ballot printing, and the randomness can be deleted after that time if no request has been received.

Note that the encrypted randomness values and symmetric key are sent directly to the printer and not posted on the WBB. As such, there is no publicly available information that could be combined with a stolen, but previously honest, printer to reveal used ballots.
The symmetric keys $sk_i$ should be deleted from the printers after the $\RT_i$ tables have been decrypted.

\subsubsection{A faster variant with a shorter permutation commitment}  \label{sec:fasterVariant}
Although the above protocol is quite efficient, it still requires the WBB to do a lot of computation to open all relevant commitments and reconstruct the ballot permutation each time a print confirmation check occurs.  This is unfortunate because we would like to encourage ordinary voters to perform print confirmations by making them easy and fast.

One way to speed up print confirming is to ask the printer to commit to the candidate-list permutation $\pi$ directly when it generates the ballot, then ensure that this commitment to $\pi$ is confirmed for proper generation (during ballot generation confirmations) and for conformance with the ballot permutation (during print confirmations).  During a print confirmation check, the WBB needs only to open and verify the commitment to $\pi$, then sign it and return it to the printer.  Every print confirmation check then triggers a ballot generation confirmation check, which opens all the commitments just as described in Section~\ref{subsubsec:ballotGenAudit}, but this does not have to be done while the voter is waiting for the print confirmation check to complete.

Like other randomness values used by the printer, the randomness used in the commitment must also be generated by the $\Mix$ servers, using a new column in each $\RT_i$ and $\CRT_i$ table.  The printer retrieves the random value the same way that it retrieves all the others, and uses it to compute the commitment to $\pi$.  That commitment is sent, along with ballot generation ciphers, to the WBB during the ballot generation stage. It is also checked, along with the ciphers, during the ballot generation confirmations so we gain a statistical assurance the commitments to the permutation are correct. During the print confirmation check everything proceeds as described above, except the WBB only has to open and check the commitment and then sign the permutation based on that. 

This does not affect universal verifiability, because the same data linking the ballot permutation to the commitments on the WBB is eventually published either way. However, the big advantage is it reduces the workload on the WBB during the critical time that the voter is waiting for the signature on $\pi$, since it can now defer the re-encryptions necessary to verify the permutation. 

More precisely, if we add the required random values into the $n+1$-th column of each $\RT_i$, Algorithm~1 would now be:
\begin{algorithm}
\center{\bf Algorithm 2: Deterministic Encryption by Printer with explicit WBB commitment to $\pi$.}
\small
\begin{algorithmic}
\For{$i = 1 \to \SC$} \Comment{Decrypt the symmetric keys from the \Mix Servers}
\State $sk_{i} \gets \dec{\psk}{esk_i}$
\EndFor
\For{$j = 1 \to \BC$} 	\Comment{$\BC$ is the number of ballots.}
\For{$k = 1 \to n$} 	\Comment{$n$ is the number of candidates.}
\State $rand \gets SHA(\aesdec{sk_1}{\RT_{1}(j,k)}\|\ldots\|\aesdec{sk_{\SC}}{\RT_{\SC}(j,k)})$
\State $CT_{j,k} \gets \enc{\epk}{\cand_k}{rand}$
\EndFor
\State $CT_{(j)} \gets Sort(CT_{(j)})$  \Comment{$CT_{(j)}$ returns the entire row}
\State $perms_j \gets \pi$	\Comment{$\pi$ is the permutation applied to sort $CT_j$.}
\State $rand2 \gets SHA(\aesdec{sk_1}{\RT_{1}(y,n+1)}\|\ldots\|\aesdec{sk_{\SC}}{\RT_{\SC}(y,n+1)})$
\State $commit_\pi \gets \com{\pi;rand2}$
\EndFor
\State Send $CT$ and $commit_\pi$  to the WBB. 
to WBB.
\end{algorithmic} \label{alg:detEncPiComm}
\end{algorithm}

Ballot generation confrimation check (Section~\ref{subsubsec:ballotGenAudit}) would be exactly as above.  Additionally, anyone can recompute the randomness value the printer used to commit to the candidate permutation $\pi$, and hence open that commitment and check that it matches the ballot permutation. 

Print confirmation (Section~\ref{subsubsec:printAudit}) would be:

\begin{description}
\item[Voter:] After collecting printed candidate list, remains at printer and requests a confirmation check,
\item[Printer:]  sends to the WBB the randomness to open the commitments to the randomness on each $\CRT_i$ used to generate the ballot in the ballot generation phase and the randomness to open the commitment to $\pi$.
\item[WBB: (immediately)] checks the serial number has not already been voted on or confirmed and if this is true, opens the commitment to $\pi$, checks it, and if valid sends a jointly signed copy of $\pi$ (or candidate names in permuted order) to the printer. 
\item[WBB: (later)]  opens all the commitments for this ballot, reconstructs the ballot, computes the permutation $\pi$, posts all the data on the public WBB.
\end{description}

The final step of opening all the other commitments reduces the total number of confirmations that need to be done.  If we relied entirely on the immediate check of the serial number and the frequency of ballot generation confirmations, then the system would still be universally verifiable, but the probability of the printer cheating successfully would be higher.

\subsubsection{Real-time Printer Replacement}
If a printer is stolen or fails, all the ballots that had been generated by it are no longer available for use. This is important because the ``printer'' is a small tablet PC that would be easy to carry. Hence we need to have a suitable method for bringing replacement equipment back online during election time. For example, we could bring the randomness generation servers back online each night, or when needed, to generate new randomness, after having deleted those values they had already sent to printers.  An alternative is for the randomness generation authorities to do the same thing as described above for a few extra as-yet-undeployed printers, put the data on the public WBB, and then ask each one to send their $sk_i$ to some (distinct) entity who is going to be online at voting time.

\section{Electronic Ballot Marker (EBM)} 
The EBM is a computer that assists the user in filling in a \pret{} ballot.  The device is already adequately described in the Introduction.  The protocol for a user to start the EBM to make it ready to receive a vote is given in Figure~\ref{fig:StartEBM}; and the protocol for casting a vote once it has been provided to the EBM is given in Figure~\ref{fig:Vote}.

\begin{figure}[th]
\begin{msc}{Start EBM}\label{msc:startevm}
\setmscvalues{large}
\setmscscale{1.0}
\declinst{vtr}{}{Voter}
\declinst{ebmt}{}{EBMTablet}
\declinst{wbb}{}{Private WBB}
\small
\nextlevel
\mess{serialNo,district,serialSig,commitTime,permutation}{vtr}{ebmt}
\nextlevel
\mess{boothID,boothSig,serialNo,serialSig,``startevm'',district}{ebmt}{wbb}
\nextlevel
\mess{serialNo,``startevm'',peerID,peerSig,commitTime}{wbb}{ebmt}
\nextlevel
\end{msc}

\begin{itemize}
  \item boothSig : {Sign$_{EBMTablet}$\{``startevm'',serialNo,district\}}
  \item serialSig : {Sign$_{WBB}$\{serialNo,district\}}
  \item peerSig : {Sign$_{WBB}$\{``startevm'',serialNo,district\}}
\end{itemize}
\caption{Start EBM Message Sequence Chart}
\label{fig:StartEBM}
\end{figure}

\begin{figure}
\begin{msc}{Vote Message Sequence Chart}\label{msc:vote}
\setmscvalues{large}
\setmscscale{1.0}
\declinst{vtr}{}{Voter}
\declinst{ebmt}{}{EBMTablet}
\declinst{wbb}{}{Private WBB}
\small

\mess{serialNo,district,serialSig,commitTime,permutation}{vtr}{ebmt}
\nextlevel
\mess{races,boothID,boothSig,serialNo,serialSig,``vote'',startEVMSig,district}{ebmt}{wbb}
\nextlevel
\mess{serialNo,``vote'',peerID,peerSig,commitTime}{wbb}{ebmt}
\nextlevel
\mess{serialNo,district,serialSig,permutation,peerID,peerSig,commitTime}{ebmt}{vtr}
\end{msc}

\begin{itemize}
  \item boothSig : {Sign$_{EBMTablet}$\{serialNo,district,preferences\}}
  \item serialSig : {Sign$_{WBB}$\{serialNo,district\}}
  \item startEVMSig : {Sign$_{WBB}$\{``startevm'',serialNo,district\}}
  \item peerSig : {Sign$_{WBB}$\{serialNo,district,preferences,commitTime\}}
\end{itemize}
\caption{Vote Message Sequence Chart}
\label{fig:Vote}
\end{figure}

\section{Cancel Station} \label{sec:voteCancel}
The Cancel Station is a supervised interface for cancelling a vote (identified by Serial Number).  This will be implemented on the same devices as the print-on-demand printers, but remains conceptually distinct (and could easily be implemented on a separate device).  

To request cancellation of a vote, a voter presents the printed candidate list to the election staff.  The candidate list contains the signed serial number, which is scanned by the cancel station and passed to the Cancel Authority to run centrally.  That cancel authority constructs its own signature on the cancellation request and sends it back to the Cancel Station.  That cancel authorisation is  then forwarded to the Bulletin Board peers for inclusion on the bulletin board.  A cancellation receipt is returned for the voter to retain, signed by a threshold of the bulletin board peers.   This protocol is illustrated in Figure~\ref{fig:posting}.
A precise description of the surrounding procedures is given  in Section~\ref{subsec:cancel}.  
It is important to emphasise that a vote is never cancelled unless the elector follows those procedures.

Each Cancel Station has  a (low) limit on the number of cancellations that may be submitted through it before being checked by VEC HQ.  Hence a malicious Cancel Station incorrectly requesting cancellation of votes will not be able to cancel many before it is under scrutiny.  Furthermore the cancel requests may be revoked if they are discovered to be flawed, and the  votes can then be reinstated.

\begin{figure}[t]
\begin{msc}{Cancellation}\label{msc:cancel}
\setmscvalues{large}
\setmscscale{1}
\declinst{vps}{}{Cancel Station}{}
\declinst{ca}{}{Cancel Authority}
\declinst{wbb}{}{Private WBB}
\small

\mess{\small serialNo, serialSig,``cancelReq'', district, senderID, senderSignature}{vps}{ca}
\nextlevel
\mess{\small CancelAuthResp}{ca}{vps}
\nextlevel
\mess{\small boothID, boothSig, cancelAuthID, cancelAuthSig, serialNo, serialSig,``cancel'', district}{vps}{wbb}
\nextlevel
\mess{\small serialNo, ``cancel'', peerID, peerSig, commitTime}{wbb}{vps}
\end{msc}

\begin{itemize}
  \item senderSignature : {Sign$_{VPS}$\{``cancelReq'',serialNo,district\}}
  \item CancelAuthResp : {Sign$_{CancelAuth}$\{cancelAuthID,cancelAuthSig\}}
  \item boothSig : {Sign$_{VPS}$\{serialNo,``cancel''\}}
  \item cancelAuthSig : {Sign$_{VPS}$\{serialNo,``cancel''\}}
  \item serialSig : {Sign$_{WBB}$\{serialNo,district\}}
  \item peerSig : {Sign$_{WBB}$\{``cancel'',serialNo\}}
\end{itemize}
\caption{Cancellation Message Sequence Chart}\label{fig:posting}
\end{figure}

\section{Mixnet} \label{sec:mix}
The system uses a re-encryption mixnet which produces a noninteractive, universally verifiable proof of a shuffle and decryption (of encrypted votes) and posts it to the Public WBB.  The main advantage of the re-encryption (as opposed to decryption) mixnet is that it separates the processes of shuffling and decryption.  If one mix server blocks, it can simply be removed and the mix re-run.  Decryption works as long as a threshold of key share holders remains available.

The current implementation uses Randomised Partial Checking \cite{jakobsson02:e-vote} with the modifications proposed by Khazaei and Wikstr\"om \cite{khazaei2013randomized},
including requiring all mixnet peers to contribute to the randomly 
generated challenges.  

Each division will be shuffled separately.

There will be a problem if the number of votes cast in a particular division
(for ATL or BTL) is too small to be an acceptable anonymity set.  This is an unavoidable problem that has nothing to do with the choice of mix protocol, though it is affected by the decision to allow receipts to reveal the choice of ATL or BTL.  This means that that data, along with the division name, will be revealed on the mix inputs, so each category needs to be mixed separately.  Some categories could be quite small.  It will be up to the administrators to decide how small an anonymity set must be before the disadvantages of privacy compromise outweigh the advantages of public verifiability.

We would pad all votes going through the mix to the same length. If someone provides 27 preferences, but everyone else provides 36, we will pad the set of ciphers for the 27 to be the same length. The padding is a zero/null value that is agreed and published in advance. That prevents trivially tracking the data through the mix. However, it does not prevent matching a receipt of 27 preferences to a decryption of 27 preferences.  See Section~\ref{subsec:complexBallots} for a discussion of the options and implications.

\VTNote{At some point one of James, Chris or Vanessa, may fill this in.  Or maybe not.  The above is just passable as is.  I've added but commented
out an appendix heading, which I think would be the right place 
for it.}

\chapter{Robustness and recovery from failures}\label{sec:robustness-recovery}

\VTNote{This possibly needs a sec on bulk cancellation, i.e. the case in which there's a good reason for believing that a whole set of votes is wrong.  e.g. when the device was stolen but the theft not noticed.  This should happen only in very rare circumstances, and never for a voter who has left the polling place with a signed receipt.  Very tempted to leave it out altogether, as CB says it is ``outside the protocol.''  SAS: leave out for now (Jan14)}

Section~\ref{sec:polling-place-procedures} described a series of checks that voters and others can perform to
ensure integrity, but did not specify exactly what happens when any of these checks fail.  It 
is challenging in any voting system to recover from errors.  A failure of 
check~1, 3 or~4 is immediately demonstrable (assuming Check 3 is performed on 
the spot in the polling place) and proves malfeasance by election authorities.  
This would have serious implications for the trustworthiness of the election 
result.  It is less clear how seriously to regard a failure of Check~2.  
Unfortunately there will be some rate of false alarms, in which voters claim 
their vote was misrecorded when they simply misremembered it or changed their 
minds.  Hence a zero-tolerance policy is unworkable, even though any tolerance 
increases the chances for vote manipulation.  Whatever the level of tolerance, 
it is important that ballots spoiled in this way remain secret, or the process 
can introduce opportunities for coercion.

One simple fallback method, in the event of unacceptable failures including 
the loss of network connectivity or complete device failure ({\it e.g.} a printer or EBM malfunctioning irretrievably), is simply to stop using \pret{} and use the EBM as a plain electronic ballot marker.

\section{Fallback to Plain EBM mode} \label{sec:plainEBM}
 The protocol requires the participation of the Private WBB.  Intermittent loss of connection to the WBB will be handled by the communication infrastructure, but loss of the network or the WBB will require fallback to a mode of operation that is purely local.

Since vVote is intended to operate alongside standard paper ballots (for 2014) the fallback position will be to return to ``plain Electronic ballot marker mode'': the EBM machines will be used purely for constructing and printing a paper ballot which can then be cast with the standard paper ballots.  This would contain the voter's preferences in boxes beside candidate names (and parties/groups) listed in the offical order, without the serial number or any other identifying information.  The result would look as much like the manually-completed paper ballots as possible given the printer available.\footnote{Legislation specifies certain formatting rules for ballots, so technically speaking these would not be ballots, but some sort of summary representation of the voter's intention that could be counted manually in place of a ballot.}

No receipts will be issued and the vote will not be submitted through {vVote}.  The Print on Demand service is not used at all for plain EBM voting.

This provides simple cast-as-intended verification, but no more evidence than the ordinary paper system that the vote was properly (transported and) included in the count.  In other words, this system is software independent but still depends on procedures in the polling place and possibly in the transport of paper votes.  This is a reasonable fallback in the event that vVote is unavailable, particularly for voters who would require assistance to complete a paper ballot manually.

Of course if there had been a complete power failure then all voters could have reverted to paper voting.

\section{Recovery from other failures}
The rest of this section describes recovery strategies from failures that are not so overwhelming as to motivate
reverting to plain EBM mode.  To date little work has been done on robustness, which mainly concerns analysing and providing safeguards against possible system faults. We define robustness of an e-voting system as the ability to tolerate failure of any of its components. On the practical level, recovery from failure/misbehaviour should also be timely. 

The importance of adequate recovery measures is highlighted when a system is deployed for the real world. Provision for occurrences such as network or power outages, device malfunction, {\it etc.}, that do not feature significantly in theoretical work becomes a more pressing requirement. Something as apparently innocent yet all too real as a printer breaking down could, for instance, bring voting to a standstill, no matter how secure the protocol is proven otherwise.

As might be expected for complex systems which require a high level of security and reliability, the task of accounting for all possible failures in an e-voting system is extremely involved, and demands rigorous attention to detail. It is crucial that any authority running an election is ready with appropriate procedures to deal with any deviations from a normal protocol that may occur: from the seemingly mundane, such as printers running out of paper to the more serious, such as errors in device digital signatures.

Officials should keep detailed records of device and procedural failures of any kind. Repeated and/or widespread occurrences, e.g. in the device failing to sign receipts may necessitate de-commissioning and replacing a device, and forensic examination to determine the possible cause. The authority may need to specify a number for each type of failure, after which further investigation is carried out.

We will walk through the vVote voting ceremony and consider the various error states that may arise, and the possible remedial procedures to ensure robustness of the system.  Many of the circumstances require a vote to be \emph{cancelled}, which has its own special set of procedures, described in Section~\ref{subsec:cancel}.

We assume there is an established procedure for registering and authenticating voters at the polling station, for example marking off the  electoral roll on production of some recognised form of identification or claim of identity. 
This may not be trivial if the register is electronic, in which case it may be affected by a network or power outage (but this is out of scope).

\section{Potential failures in ballot generation or ballot printing confirmation}
The authority should encourage voters and third parties who may wish to confirm ballots during either the ballot construction phase or the voting phase.

\begin{description}
\item[Print station:]
 does not work, either for printing or for confirming, or does not print a valid WBB signature on the ballot:

Persistent device failure has been discussed in Section~\ref{sec:plainEBM}.

A printer that occasionally fails to produce a proof of correct opening in response to a confirmation request should be treated with suspicion.  In the absence of a good reason for expecting network failures (preventing the print station from contacting the private WBB), the assumption should be that the ballot is not properly formed.  The threshold for replacing such a printer should be low.  A printer that had printed a
malformed ballot would not necessarily fail by proving that the ballot was malformed---more likely, it would fail to provide a proof at all.

\item[Ballot is incorrectly formed: ]

Finding inconsistencies in ballot formation could potentially occur during pre-election checking, or during voter-initiated checking during the voting phase. 

A printed ballot with a proper WBB signature but a candidate list that is inconsistent with the cryptographic information opened in a confirmation check, represents a serious failure of the system.  The affected device should not be used.  \VTNote{Check if this calls the WBB into question.  SAS: I don't think it does, the WBB only signs the serial number}

\end{description}

\section{Potential failures in the voting ceremony}~\label{vc-errors}

After obtaining a printed ballot for voting, the voter interacts with an EBM and possibly a device that checks and verifies the WBB signatures. Potential failures and possible recovery measures are discussed below.  There are several different potential failures, but they may be difficult to distinguish and hence have similar responses.

\subsection{Demonstrable EBM malfunctions}
\begin{description}
\item[The preference receipt is invalid] For example, it is blank or it contains repeated numbers.  
This is an EBM error.
\item[EBM receipt is printed but does not contain a valid digital signature: ]
This is also an EBM error.  An invalid or absent signature may be the result of simple device malfunction, but it may also have more serious implications such as failure to upload the vote to the WBB,  or incorrect recording on the WBB. 
\item[Recommended action in both cases:] \ \\[-3ex]
	\begin{itemize}
	\item The vote should be cancelled, then
	\item The voter should be issued with a fresh ballot and allowed to cast another vote. 
	\item The EBM should be removed from use until the problem is resolved.
	\end{itemize}
\end{description}

 These steps should be auditable, {\it i.e.} it should be possible for voters and third parties to check cancelled votes on the WBB.   See Section~\ref{sec:voteCancel} for details of procedures.

\subsection{Apparent malfunctions that may be due to voter, EBM, or private WBB error}
In all of these cases, it is possible that the EBM malfunctioned but also possible that a voter did not follow the instructions correctly, or claimed that the EBM malfunctioned when it did not.
\begin{description}
\item[The EBM does not print a preference receipt] 
The elector cannot know if the VEC captured their vote as they intended it because they cannot verify it.
\item[The preferences receipt is different to what the elector expects]
Note that this is indistinguishable from an elector who is mistaken, frustrated or dissatisfied claiming that their receipt is different when in fact the device functioned correctly.
\item[Recommended action in both cases:] \ \\[-3ex]
	\begin{itemize}
	\item The vote should be cancelled, then
	\item The voter should be issued with a fresh ballot and allowed to cast another vote. 
	\end{itemize}
\end{description}
It is not possible to tell whether the failure was due to the EBM failing to send the correct vote, or a network failure, or the private WBB failing to sign and return what it received.  Vote cancellation and re-voting are the most appropriate actions as the system may have failed to record the vote.  However, it would be premature to remove the EBM from service following only a small number of accusations of this sort of behaviour.  
The authority may decide to treat sporadic, or a small number of errors as simple anomalies or voter mistakes.    However, the authority should investigate the cause of a large number of reported errors. 

\section{Potential failures in post-election checking}~\label{audit-errors}

\begin{description}
\item[Vote is incorrectly recorded on the WBB or absent from WBB: ]
On discovering such an error, a voter can lodge a complaint. 
The first step would be for an official to check the reported error and the WBB signature on the receipt. If the vote is indeed wrongly recorded, and the voter has a validly signed receipt, this represents a serious system failure which implies that more than the assumed maximum number of private WBB peers have been compromised.

A voter who makes this complaint without a validly signed receipt does not have a strong case.

\commentOut{
It would be wise to implement a paper audit trail~\cite{mercuri:e-vote},~\cite{vepat:e-vote} as a fall-back in case of technical failure of the system.  A paper audit trail would provide a fall-back if there is a fault in the ``back-end'' processing which cannot be quickly fixed. }

\item[Final tally proof does not verify: ] 

An error found post-publication is more likely to be caused by a technical problem e.g., in uploading, than in the actual calculation. 
Clearly, the error would have to be investigated, the problem located and a solution sought. 
While it is potentially very serious, possibly indicating failure in the ``back-end'' processing, it could also be the result of a minor error that is easily fixed by, for example, uploading missing data.

If no simple error is found, it should be possible to identify the mix servers whose proofs were not valid (or not present) and rerun the mix without them, eventually producing a valid, verifiable proof.
\end{description}

\chapter{Security Claims and analysis}\label{sec:security-analysis}
Security requirements for voting systems fit into two main categories: integrity properties and privacy properties.  We give a brief overview here and then more detail on each set of properties.

For sighted voters, the protocol includes no human or electronic components which must be trusted for integrity, apart from
trusting that each eligible voter is allowed to cast at most one vote,
and that only eligible voters can vote.  It does of course rely on
voters to perform some checks (an auditing assumption), which are detailed in
Section~\ref{sec:polling-place-procedures}. It assumes that they can find at least one honest device to view the bulletin board, check signatures and verify ballot confirmations.  
 Invalid ballots, in which the candidate
list does not match the encrypted ciphertexts on the WBB, are detected at ballot confirmation by
Check~1.  Check~2 detects incorrect vote printing by the EBM.  Incorrect vote 
submission by the EBM before submission to the private WBB is detected by
Check~3.  Check~4 detects vote substitution by the WBB.  Incorrect
mixing or decryption would be detected because the proofs of correct
mixing and decryption are public.

The vision-impaired voter is unable to do Check~2, that the EBM
printed the correct ballot. She cannot ask for human assistance without
destroying privacy.  This leads to a distribution of trust over the machines
in the polling place: she can check her vote on as many machines as she likes,
and must assume that at least one of the machines she uses is honest.

Some vision impaired voters have good enough vision to check their printout directly, just like ordinary voters, without using a second EBM.  The harder it is for a cheating EBM to predict who will check directly, the harder it is to get away with cheating.

The protocol not only detects, but also provides evidence of many kinds of failures, including the failure of ballot generation confirmations and the failure of a properly produced receipt to appear on the bulletin board.  In both cases, a voter who experiences such a failure has a WBB digital signature that proves that the system malfunctioned.  This provides two kinds of benefits: voters can demonstrate to a court that a real malfunction occurred, but it is not feasible to pretend that a malfunction occured when it did not.  This is important in defending against the ``defaming attack'' in which people pretend to have detected a system failure which did not actually happen. Of course, there can be no proof that the EBM accurately represented the voter's intention: that step is dependent on the voter's testimony and hence is to an extent vulnerable to the ``defaming'' attack.

Privacy of the contents of each receipt depends on the assumption that at least two \Mix{} Server
generate randomness correctly and keep it secret. Further,
that a threshold set of those who share the keys is honest.  If these assumptions hold,
then the receipt itself does not leak information about the voter's preferred candidates 
(though it does show how many preferences they listed, and whether they voted ATL or BTL).

Provided that that two assumptions holds, the system has some defence
against kleptographic attacks on the
receipt~\cite{DBLP:conf/etrics/GogolewskiKKKLZ06}.  This is because the
receipt's random data is generated in a distributed way, and the entities that
do the printing (the printer and the EBM) are deterministic.  Thus information
cannot be leaked in the ballot data itself without some chance of detection, 
though it could be subtly leaked
in slight font changes or other printing effects.

Privacy of the votes also depends on the privacy of the mixing protocol.  
If the mix is secure, then the tallying protocol does not add any
information about the link between a receipt and its vote.  RPC challenges are constructed so that overall anonymity across the mix is preserved.  
\VTNote{More precise claim about RPC?:  SAS: a bit more explicit, not sure what claim you want.}

The system is also receipt-free, meaning that it does not provide a person with information to prove how she voted.
However, there are coercion attacks on this protocol, including the ``Italian
attack.''  These, along with
other important details, are described below.  We concentrate first on 
integrity properties and then discuss the subtleties of privacy.
\VTNote{RF claim doesn't quite square with presence of coercion attacks.  Perhaps we're saying that the protocol doesn't add 
any coercion attacks that aren't already present in the system?  Is that true?   SAS: half true: VEC already publish the votes, so Italian attacks are already possible, and so we don't add this capability to the attacker, because she already has it.  On the other hand, if VEC didn't publish the votes already then we would be introducing it because we require that the votes are published for e2e}

This protocol is meant to achieve two main classes of security properties: integrity and privacy.  The claims are given here and their informal justification is given below.

\section{Integrity properties} \label{sec:security-integrity}
\begin{description}
\item[vote integrity: ] meaning that all attempts to manipulate the votes are detectable by confirmation checks or other audits.\footnote{Of course this does not imply that they will always be detected, if the appropriate checks are not performed on the manipulated ballot.  The claim is that any manipulation can in principle be detected if a check is performed.}  
\item[non-repudiation: ] meaning that failures can not only be detected, but (in most cases) demonstrated.  In particular, failures of the private WBB to post something it has accepted on the public WBB can be proven by producing the signed accepted item (whether a signed ballot confirmation or a signed receipt for a submitted vote).  This also defends the system against people falsely claiming to have detected an error.
\item[prevention of ballot stuffing: ] meaning that (under a threshold assumption, and a procedural assumption for voter markoff) only votes entered via the legitimate interface are included in the count.    

Defences against ballot stuffing using the legitimate interface (e.g. by unauthorised people gaining access to a legitimate polling place) are not part of the system and must be defended against by procedural mechanisms.
\end{description}

Procedures must prevent voters from taking someone else's ballot off the printer and hence voting in the wrong division.

\subsection{Justification of Integrity claims} \label{sec:Integrity Analysis}
\VTNote{Add discussion of mixes and their integrity properties}
\subsubsection{Vote integrity based on confirmations}
An informal argument for the integrity of each person's vote is:

\begin{itemize}
\item The ballot-generation confirmation checks that the ballot is a permutation of properly-encrypted candidate identifiers.
\item The ballot-printing confirmation checks that the printed list of candidate names matches the encrypted candidate identifiers on the WBB.
\item The voter's check of the EBM's printout confirms that the correct numbers (or other marks) are recorded against the correct candidate names.
\item The signature check confirms that the printed number sequence matches what was submitted to the WBB.
\item The check of the vote on the WBB confirms that the correct ciphertexts were used (in the case of a larger-than-necessary generated ballot) and that the vote submitted to the WBB was posted.

\item verifying the shuffling and decryption proof from the mixnet confirms that the announced output votes match those posted (encrypted) on the WBB.
\end{itemize}

Of course the first two confirmations are performed only on ballots that are \emph{not} subsequently voted on.  The argument is that any attempt to manipulate the vote by generating or printing invalid votes will be detected by with some probability that depends on the confirmations being numerous and unpredictable.

\subsection{Selecting Ballots for Confirmations and Audits} \label{subSec:auditCount}
Clearly it is important that we use a suitable source of randomness for the selection of the ballots to do the confirmation checks.  Some combination of 
public confirming (with officials, scrutineers, observers, and a public source of randomness such as dice or lotto balls) with 
voter-initiated confirmation checks (in which any voter may choose to confirm ballot construction or printing) would be ideal.  This will require further investigation to see what procedures are possible in practice.  The argument about the integrity of the results of the election depends on
all steps of the confirming process having been performed diligently, 
including those that the voters have to do themselves.

Of course, it is difficult to compute the appropriate amount of confirming for an IRV/STV election, especially in advance \cite{Magrino:irv,Cary:irv}.  This question will have to be addressed for the project, but is out of scope for this paper.  We expect that most of the IRV ({\it i.e. } single-seat) contests will have a relatively easy margin computation in practice.  However, the full STV contests are a different matter and require significant further thought.  One possibility is to announce the result, explain what quantity of cheating might have been possible given reasonable estimates of the amount of confirming that was done, and ask any election challenger to demonstrate a set of votes in which they win a seat and the number of changed votes is reasonably probable given the rate of checking.\footnote{Thanks to Ron Rivest for this suggestion.}

A Bayesian method of bounding the total number of errors for a particular confidence level, given a particular quantity of confirming, is given in Appendix~\ref{app:BayesErrorCount}. 

\subsection{Properties of different kinds of mixnet}
This design is largely independent of the kind of mixnet chosen to shuffle and decrypt the votes.  However, it inherits the privacy properties of the mixnet it uses.  The current implementation uses Randomised Partial Checking \cite{jakobsson02:e-vote}, adapted according to recommendations by Khazaei and Wikstr\"om \cite{khazaei2013randomized}.  As described in Section~\ref{sec:mix}, RPC mixnets allow a small but non-negligible probability of successful cheating, which decreases exponentially with the number of substituted votes.  

Our implementation is designed to allow cheating only if the entire set of mixers collude to cheat, because they all work together to compute their challenges.  Even if they all do, a given probability of detection requires an amount of precomputation work that is exponential in the number of substituted votes.

We intend to continue to evaluate whether a mixnet based on zero knowledge proofs such as Verificatum \cite{Wik10} will be feasible to use in future.

\subsection{Other integrity properties}
\subsubsection{Serial Number uniqueness and defence against the ``clash attack''}
The ``clash attack'' \cite{KuestersTruderungVogt-SP-2012} is a vote dropping technique that applies to many cryptographic voting schemes.  An attacker (as a server or ballot generator) arranges to give several different voters identical receipts.  All affected voters see their receipt appear on the public WBB, and yet only one vote has been counted.  In our protocol, the serial numbers are carefully generated to guarantee their uniqueness (See Section~\ref{subsubsec:preBallotGen}), but this doesn't prevent a corrupt printer from printing off exactly the same ballot, with the same Serial Number, for many different voters.  The printer would have to collude with a corrupt EBM that merely reused the private WBB signature, without resubmitting multiple instances of the same vote to the WBB.  There are two reasons that this represents an acceptable risk. 
\begin{itemize}
\item The attack works only if the voters subsequently cast identical votes---otherwise the cheating EBM will be unable to produce a valid signature on the receipt, and unable to post it to the WBB.  (The attack is in general harder for \pret{} than for direct-encrypting schemes such as Helios and Wombat, because the attacker must commit to the identical ballot before learning the person's vote.)
\item The attack is detectable by ballot printing audit, which would fail because the ballot has already been voted on.
\end{itemize}
Overall this attack is no more effective, requires more conspirators, and has a higher probability of detection than the simple misalignment of the candidate names on the ballot by a corrupt printer.  Hence it is appropriate ({\it i.e.} conservative) to include this attack implicitly in computations quantifying the extent of ballot misprinting, without explicitly counting it. 

\subsubsection{Non-repudiation and defence against the ``defaming attack''}
Every legitimate receipt includes a WBB signature on the SerialNumber, preferences and division.  Voters are encouraged to check the signature before leaving the polling place.  If a voter can produce such a receipt without it appearing on the WBB, then this demonstrates that the Private WBB has malfunctioned.  Conversely, accusations that a particular receipt was properly submitted remain unconvincing when the claimed receipt does not have a valid WBB signature.

Ballot generation and printing confirmations can demonstrably fail, or demonstrably succeed, or fail silently ({\it e.g.} when the device being confirmed simply stops).  In other words, if a voter claims to have received an invalid opening of a printed ballot, then that should be demonstrable because a printed ballot should have a valid WBB signature.  However, if a voter claims to have attempted to confirm a ballot but not received a result, this cannot be checked.

Misprinting of voter intention by the EBM cannot be demonstrated, because only the voter knows what they truly entered.  Consequently, accusations of incorrect printing cannot be repudiated---the evidence that a particular machine is misbehaving needs to consist of a series of observations and comparisons with other machines.

\subsubsection{Prevention of ballot stuffing}
The private WBB enforces that votes may only be uploaded to the WBB by a legitimate EBM casting a vote that has been properly printed and signed.  In other words, a colluding printer and EBM could stuff the ballot, but this would be detected by the reconciling of markoff data with the number of submitted ballots described in Section~\ref{sec:polling-place-procedures}.  Also (a threshold of peers of) the private WBB could stuff the ballot, by pretending to have received a legitimate signed vote, but again this is detected by reconciling with markoff data.

It is not possible to include the EBM signature on the public WBB because the protocol does not guarantee that the same signature has been sent to all private WBB peers.

\VTNote{Is that right?  It seems a little strange now that I've written it out.  Why can't we tell the EBM to send the same sig to everyone?  Otherwise the private WBB peers can stuff the ballot alone.  Not disastrous but not ideal. If they have to include the EBM sig then they need a colluding EBM to stuff the ballot.}
 
\section{Privacy properties} \label{sec:security-privacy}
\begin{description}
\item[privacy] The system hides how each person voted, under assumptions stated below.

It does reveal whether the person voted above or below the line, and how many preferences they expressed.  This could potentially be used to coerce certain types of voting (such as a vote for a particular number of preferences), but it would not be possible to coerce a particular political effect.   Although this could have been avoided, the opportunities for coercion are very limited, and hence did not justify the extra difficulty for voters of a more complicated protocol that kept it secret.  

\item[receipt freeness] Even a voter who deliberately colludes with a coercer cannot prove after voting how they voted (except by major violations of enforced procedures, such as taking a photo of their candidate list).  Note that this implies that a voted ballot cannot also be confirmed.
\item[resistance to kleptographic attacks] A printer attempting to leak information via the WBB data will be detected with some probability.
\end{description}

It is not intended to defend against ``Italian attacks'' (in which the voter is coerced into producing a vote matching a detectable pattern) or randomisation attacks (in which a voter is coerced to produce a receipt of a particular form, which has a random effect on the actual vote).  The coercion attack described in Kelsey {\it et al. } \cite{DBLP:conf/wote/KelseyRMC10} Sec 4.3 is also possible, but so complicated for this scheme that it would be very far from the most effective way to coerce voters.
It is not worth defending against a coercion attack that is harder to execute and no more effective than the ``Italian attack'' already possible in the existing paper system.

Defences against coercion associated with failing to shred the candidate list must be enforced by procedural mechanisms.  Similarly, deliberate uses of out-of-band recording technology (such as taking a photo of the candidate list before shredding it) or side-channel information leakage (from the EBM or printer) must be defended against by mechanisms outside the vVote system.

vVote also contains some technical and procedural measures, which are unrelated to \pret{}, for defending against chain voting.  These are described in Section~\ref{subsec:chain}.

\subsection{Justification of privacy claims} \label{subsec:privacyAnalysis}
The following coalitions can violate ballot privacy for an individual:
\begin{itemize}
\item The printer that generated and produced that ballot,
\item All but one of the $\Mix{}$ servers,
\item The EBM the voter used,
\item However many mix servers are necessary for breaking shuffling privacy, depending on the mixnet being used, 
\item A threshold of key sharing authorities,
\end{itemize}
\VTNote{Nitpick: For key sharing, is the threshold the minimum number to know the key or the max to not know it?}

The crucial claim is that smaller coalitions cannot.  This is expanded into
several specific claims below.

Clearly if the printer leaks its information it can violate vote privacy for everyone who used a ballot it printed.  This means that practical opportunities for compromising the printer must be reduced as much as possible, {\it e.g.}  turning off the wireless connection.

Apart from the printer and an electronic ballot marker (if there is one), no other single entity can violate vote privacy.  This is justified in two claims below.

\begin{claim}
A collusion of all but two randomness generation authorities does not have sufficient information to recover the ballot permutation (in polynomial time with non-negligible probability).
\end{claim}

Clearly if all the randomness generation authorities collude and share their information, they learn the contents of all ballots.  If at least two choose their random values correctly and keep them secret until the others have committed, and if the others can be forced to open their commitments, the resulting list of random values has $2 \kparam = 2*256$ bits of entropy.

NIST \cite{NIST} states that the SHA family of hashes are suitable as randomness extractors. In \cite{NIST} it states that when using a hash function $F$ in which $Y=F(S||A)$ then ``If the input string S was assessed at $2n$ bits of min-entropy or more (i.e., $m \ge 2n$), then $Y$ may be considered to have $n$ bits of full entropy output''. The value of $A$ can be anything, including null, it is just additional data. This tells us that provided at least two mix servers provide good randomness values the output from the hashes will have $\kparam = 256$ bits of entropy.  

The usual subtlety arises if we consider the possibility that some authorities might use blocking to bias the output, {\it i.e.} might wait until learning the other authorities' random values and refuse to open their own commitments if they did not like the result.  (This could happen, for instance, in collusion with a corrupt printer.)  This is why true coin-tossing protocols are more complicated than the simple one in this proposal.  In practice, such a blocking authority would be removed quickly without having the opportunity to affect many bits of the output.  \VTNote{See Olivier P's paper on good enough privacy for Helios, ie. why it's OK to use Pedersen for Helios keygen even though the same issue exists.}

Clearly the same argument holds for the PRNG construction of Section~\ref{subsubsec:PRNG}, given appropriate assumptions about the PRNG.

\begin{claim}
The posted ballots on the WBB reveal, for each receipt, whether
the vote was ATL or BTL and how many preferences were cast, but no
other information, unless a threshold of key sharers colludes.
\end{claim}

\begin{claim}
The mixing process anonymises votes within anonymity sets defined
by their division, ATL/BTL choice, and length of preference list.  The
assumptions about mixnet collusion for privacy violations depend on
the mixnet being used.
\end{claim}
For example, in RPC, if all but one pair of mix servers exposes their permutation, then each anonymity set is half the total being mixed.

In a mixnet based on zero knowledge proofs, votes are anonymised within the whole set if at least one mix server remains honest.

Our system is not susceptible to the replay attack
described in \cite{khazaei2013randomized} because all ballots are pregenerated in a distributed fashion.

\subsection{Kleptographic attacks}

The output of the printers is entirely determined by the randomness that is sent to them, and other publicly committed information given in Figures~\ref{tbl:CandIDs} and \ref{tbl:InitialBallotInput}.  Hence they have no opportunity to provide any information which may be skewed in a particular way.  Correct information posted  therefore cannot leak information from the printer.  Incorrect information will be detected with some non-negligible probability by the ballot-generation confirmation processes.

Although the whole group of randomness generation authorities can collude to mount a kleptographic attack, a similar argument to that for vote privacy shows that a smaller collusion has insufficient information.

\subsection{Receipt freeness}
Receipt freeness is a subtle property and we do not have a formal argument for it.  However, the main idea behind \pret{} is to provide the voter with either a proof of the contents of a ballot's encrypted values, or an opportunity to vote on the ballot, but never both for the same ballot.  In other words, a ballot that's allowed to be voted on should never have revealed the random values used to produce it.

The threat of using the confirmation process to expose the contents of a ballot
that has been voted on is ameliorated by the electronic locking process
described in Sec~\ref{subsubsec:printAudit}.

A voter could attempt to collude with a corrupt printer to produce a receipt, and could promise not to perform a ballot printing confirmation check, but the incorrect formation of the ballot necessary to produce such a receipt would be caught by a ballot construction confirmation check with some probability.

\subsection{Privacy Threats Ameliorated By Procedural Controls}

As the voter inputs her choices into the EBM, the device necessarily ``learns'' 
how she voted. The potential for the EBM to leak vote information clearly 
raises privacy issues.  Any data stored in the EBM's 
memory should be deleted, ideally after each session.

\pret{} introduces a privacy threat that does not exist for either standard
paper voting or for DRE's with VVPAT: someone may
discover and record an unvoted ballot's candidate order and look up code, then
learn the vote choices when they are later posted on the WBB.  Therefore there
should be procedural controls to protect both the paper printout and the
electronic data on the printer from observation by anyone but the voter.

As for any voting system, computerised or paper-based, voters may ask 
for assistance at a point that potentially violates their privacy simply because
 the assistant sees what the voter has already written or entered.  This threat 
to privacy however, exists  in the current system.

\subsection{Privacy issues arising from small populations and complex ballots}  \label{subsec:complexBallots}
Sometimes vote privacy is unachieveable: if everyone in one polling place
votes the same way, and results for that polling place are observed or announced alone, then vote privacy is not possible.  It is possible that 
some divisions will have very few vVote voters.

The small populations exacerbate existing problems caused by the 
complexity of Victorian ballots.  It is easy for a voter (or a 
coercer) to choose a BTL vote that is highly likely to be unique.
If the BTL votes are made public, this allows a voter to prove how
they voted.

Since vVote receipts expose whether the person voted ATL or BTL, and how many preferences they expressed, this problem is exacerbated again: it may
be possible to identify a vote uniquely based on its division and 
number of BTL preferences.   If only one person cast a particular number of preferences below the line in a 
division, that person's vote would have to be withheld from publicly verifiable decryption.  

Some techniques exist for ameliorating some of these problems, such
as proposals for verifiable privacy-preserving tallying using (more)
mixing and homomorphic sums \cite{shuffle-sum, heather07:e-vote}.
However, these may not be efficient enough to use in practice.  More
importantly, they work only on a complete list of votes, while the
vVote votes need to be input into an (unencrypted) existing VEC 
counting system.

This leaves us with some ad hoc approaches that are highly dependent
on how many people vote in each division, how close the election result
is, and what anonymity thresholds are deemed acceptable.  The problem could be mitigated by doing on-demand decryption of the packed candidate IDs, so we only decrypt the next pack when the previous one has been eliminated. However, that still does not guarantee that privacy will be preserved, since if all preferences are counted the full information will be made public anyway. 

\subsection{Other possible attacks}

``Psychological'' attacks are a potential threat. As an example,  a coercer manages to convince voters that he 
is able to decrypt their receipts and find out how they voted~\cite{ryan05:e-vote}.
Voter education could mitigate this attack; however psychological attacks will be a problem for virtually any end-to-end verifiable system.
\section{Conclusion, report on the deployment, and future work}
For the vVote system we have taken the original design concept of \PaV{}, and extended and customised it to the requirements of the State of Victoria, Australia.  This practical deployment entailed a whole suite of technical and procedural problems not previously addressed in the academic literature.
We have developed novel solutions to make the design applicable to the particular aspects of the target election while maintaining end-to-end verifiability, notably: the introduction of electronic ballot markers for capturing and casting votes; the requirements of preference voting whereby candidates are ranked in order of preference rather than simply selected; the design of a secure and robust web bulletin board; the requirement to print ballot papers on demand at polling places while preserving the assurances that their cryptographic construction is correct; and designing the system around procedures that are straightforward for voters and pollworkers to follow.  

This paper has described the general background and motivation for the system, and its human processes and cryptographic protocols, motivated how these tie in to other design decisions.  Finally the paper has considered the system from a robustness point of view, and has provided a security analysis of the system with particular emphasis on the privacy and integrity provided by the system.  

\subsection{Report on the deployment}
The system successfully took 1121 votes.
During the deployment, six e-votes were cancelled.  This seems to have been due to networking problems.  All the affected voters subsequently cast a paper ballot.

Many divisions had too few e-votes for publication, and hence were not verifiable.  This is a result of limited deployment, not a protocol flaw, and would disappear if more people used the system.


This project ran on a constrained budget on a very firm timetable dictated by the election cycle.  Certain tradeoffs were forced by timing, and by the current availability of certain tools, but do not necessarily represent the best in the long term.  For example, the unified scanner and EBM could be revisited with better tools for printing and better optical character recognition on the scanner.  Also the choice of mixnet could be reviewed if a more efficient zero-knowledge open source mix becomes available.

Numbers were undetermined until the last minute and so the system was developed to handle a much larger number of voters than it eventually actually did.  Many of the design tradeoffs and developments meant for larger numbers, such as the agreeement algorithm for the private WBB, were not necessary for the numbers we actually had.  

\subsection{Reflections on putting it into practice, with suggestions for future improvements}
\subsubsection{Usability}
There is a tension between verifiability, privacy and usability. For professional administrators who actually have to run an election, usability is the overwhelming priority.  This includes usability for voters and for the temporary staff hired to conduct the election on the day.  In practice any working system has to be usable.  Although we had put significant effort into a design that was easy for voters to use, the polling-place procedures need to be simpler if this is to be more widely deployed.  The challenge is to retain the usability, while also providing verifiability and privacy.

There is an important link between verifiability, usability, and the practicalities of organising a polling place.  For example, the electoral commission was able to conduct quite extensive ballot construction audits, because this could be performed before polls opened, at a time when there was not enormous pressure from other tasks.  However, ballot printing confirmations were much more problematic, requiring a voter to pause just at the critical point when they were about to go and vote.  An important item of future work is to redesign the ballot printing confirmations so as to make them easier to perform without getting in the way of the main voting process.

The protocol uses signature checking to achieve non-repudiation in many circumstances, particularly to allow voters to prove that their receipts are properly accepted by the private WBB.  In practice signature checking is complicated, both the procedures in the polling place and the prior work necessary to download the app.  Alternative weaker but simpler methods of preventing the defaming attack, such as anti-counterfeiting paper, or franking or stamping of receipts in the polling place, might offer better usability for only slightly reduced security.  

\subsubsection{Vision impaired voters and assumptions}
If the system is used more widely, this would reduce the importance of the current assumption that vision impaired voters can find a non-colluding EBM in the same polling place as the one they used to vote.  The original intention was that the system would be used for a wide class of voters, including disabled ones, and that the effect of some able-bodied voters conducting audits would thus be shared by those who were unable to conduct them themselves.  However, this first deployment was not extended to non-disabled voters within the state, which left us with the choice between inviting such voters to verify with their own device (a serious practical risk to privacy) or to rely on another device in the same polling place (which is far from ideal) or not at all (and rely on being indistinguishable from sighted voters).  It was decided to rely on the second device in the polling place, mainly because this is a harmless and additive assumption.  
Furthermore, even in this deployment some eligible voters were perfectly able to read their own printout---the possibility of some such voters checking their printout directly might deter an attack by colluding EBMs in the same polling place.

\subsection*{Acknowledgements}
\addcontentsline{toc}{chapter}{Acknowledgements}
We are grateful to the following for contributions and discussions during the formulation of this design:  
Craig Burton, Matthew Casey, James Heather, Rui Joaquim,  Thea Peacock, Olivier Pereira, Sriramkrishnan Srinivasan, Roland Wen, Jason White, Douglas Wikstr\"om,  Zhe (Joson) Xia and Karen Young.   We are also grateful to Ron Rivest for the suggestion in Section~\ref{subSec:auditCount}, and to the anonyous reviewers for their careful and extensive comments and suggestions on this paper.

The name ``vVote'' is a trademark of the Victorian Electoral Commission.

This work was supported by the EPSRC Trustworthy Voting Systems project EP/G025797/1, and by the Fonds National de Recherche (FNR) Luxembourg  SeRTVS project.

\bibliographystyle{alpha}
\bibliography{e-vote}

\appendix
\chapter{Audit Sample Sizes and Confidence Levels: a Bayesian Analysis}  \label{app:BayesErrorCount}

In vVote there are several stages of processing the ballots and votes which are checked by sampling: ballot generation, ballot printing, printing of preferences, posting the preferences on the web bulletin board, and mixing by the re-encryption mixnet.   This section considers the rate of sampling necessary at each of these stages in order to make statements about the probability of there having been an undetected attack altering sufficient ballots to change the result.  

Stages are checked by sampling: selecting some of the ballots/votes being processed, and checking that they have been processed correctly.  If a ballot passes a check then it has not been altered.  The stages to check are as follows:
\begin{description}
\item[Ballot generation:] Cut and choose: ballots are sampled and checked that the candidate identities have been encrypted correctly.   Sampled ballots are not used for voting, since the confirmation check would break secrecy of the ballot.  This check is carried out by the authorities during the ballot generation phase.
\item[Ballot printing:] Cut and choose: printed ballots can be challenged by a voter, who can have it checked that the printed candidate list has been correctly printed.  Checked printed ballots are not used for voting since secrecy would be lost.  This check is carried out by the individual voters.
\item[Printing of preferences:] Voters check that the preferences on their signed printed receipt are in the correct order.   Carried out by individual voters.
\item[Checking the WBB:] Voters check that the entry on the WBB matches their receipt.  Carried out by individual voters.
\item[Mixnet:] Randomised partial checking of the links in the mixnet.  Carried out by the authorities.
\end{description}

We are interested in the claims we can make when all of these checks are successful.  

The kind of claim we wish to make is that {\bf given a sampling rate of $r$ and no failed checks, the probability that at least $F$ votes were altered is no more than $p$}.  Typically $F$ will be some small fraction of the total number of votes cast.  The value $F$ of interest may be the smallest number of votes necessary to change the result.  

The converse is straightforward to calculate:  if $F$ votes are altered, then the probability that this is not detected by sampling at rate $r$ will be $(1-r)^F$ (see Appendix~\ref{app:BA}).  We can then use Bayes' theorem to obtain the probability that there is an attack given that the sampling does not find any cheating, as follows: 

Given a prior (pre-sampling) probability $q$ that there has been an attack affecting $F$ ballots, we can calculate the probability $p$ that there has been an attack affecting $F$ ballots given the sampling.  This is given by the following Bayesian formula, which is explained more fully in Section~\ref{app:bayes}:
\begin{eqnarray}
p & = & \frac{(1-r)^F .\; q}{(1-r)^F .\; q + (1-q)}  \label{bayes}
\end{eqnarray}

\section{Digression: combining confirmations of different stages}

Given stages $1$ to $n$, to change a total of $F$ votes an adversary may change some votes at each stage, to a total of $F$.  We consider $F_i$ votes changed at stage $i$, where $F_1 + \ldots + F_n = F$.

Each stage may have a different rate of sampling.  At stage $i$ we consider the rate of sampling to be $r_i$. The probability that the sampling at stage $i$ will not find any of the $F_i$ altered ballots is $(1-r_i)^{F_i}$.  Hence the overall probability $p_0$ that none of the attacked ballots will be discovered will be $p_0 = (1-r_1)^{F_1} . (1-r_2)^{F_2} . \ldots . (1 - r_n)^{F_n}$.  

Let $r_J$ be the least of all the rates.  Then (replacing each $r_i$ by $r_J$), $p$ is bounded above by $(1-r_J)^F$.  Hence we have the highest probability that none of the attacked ballots will be discovered when we make all the changes in the stage with the lowest sampling rate.  

Hence when we identify a sampling rate required to achieve particular assurances, we require that the identified rate will be the {\em minimum} that any of the stages have.  
This means that in {\em all} the stages of ballot processing which involve auditing, the sampling must be at the level of {\em at least} at the required rate.   

Note that the RPC check of the mixnet requires a different argument because it is links in the mix rather than individual ballots that are checked.  In practice the best known attack on mixnet integrity allows a vote to be altered with a $25\%$ probability of detection, and can thus be treated as a $25\%$ rate of sampling.  In practice this will be higher than the sampling in the other stages.


%
%

\begin{figure} 
\begin{center}
\includegraphics[width=\textwidth]{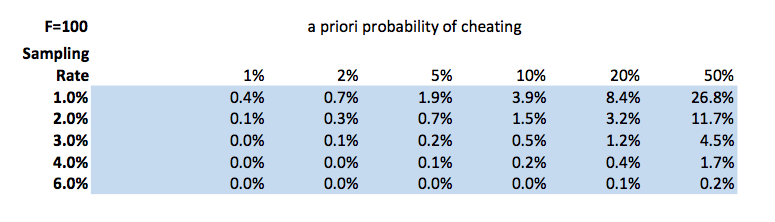}
\end{center}
\caption{Calculations of probability of fraud given no fraudulent ballots in sample} \label{fig:calcs}
\end{figure}

\subsection{Example}

Figure~\ref{fig:calcs} gives an example table, considering the probability for $F = 100$ of fraudulent ballots given a percentage sample size which finds no fraudulent ballots, and for various a priori probabilities of attack.  

For example, suppose that we have an a priori expectation from our other information (e.g. no other circumstantial evidence of any attack behaviour) that there is at most a $5\%$ chance of an attack, so $q = 0.05$.  Suppose also that $r = 3\%$ and that $F =100$.  Then we have $(1-r)^F = 0.05$, and we obtain $p = 0.002$.  

In other words, if we sample $3\%$ of the ballots, and we find no fraudulent ones, and we also estimate that there is no more than a $5\%$ probability of an attack, then the probability of there being an attack on 100 ballots given the sampling is $0.002 = 0.2\%$.

If the prior probability of an attack was at the higher value of $10\%$ then after the sampling the probability of there having been an attack will be $0.5\%$.  

\subsection{Required sampling rates}

\begin{figure} 
\begin{center}
\includegraphics[width=\textwidth]{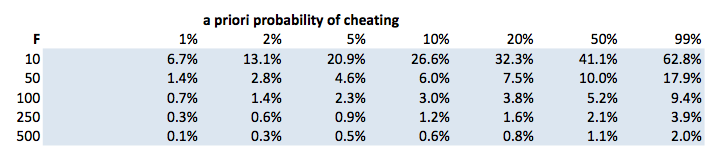}
\end{center}
\caption{Sampling rates required to achieve {\bf 99.5\%} probability that that there was no cheating to the level of $F$ altered ballots.} \label{fig:samps}
\end{figure}

\begin{figure} 
\begin{center}
\includegraphics[width=\textwidth]{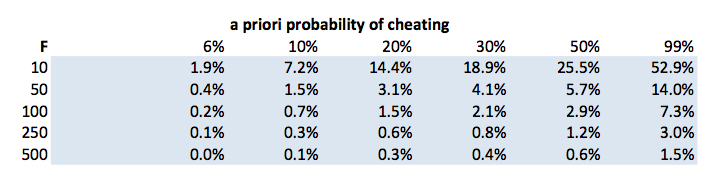}
\end{center}
\caption{Sampling rates required to achieve {\bf 95\%} probability that that there was no cheating to the level of $F$ altered ballots.} \label{fig:samps95}
\end{figure}

In practice we will start with a number $F$ of fraudulent ballots that we want to make a claim about; a judgement on the prior probability $q$ of an attack; and a confidence level $1-p$ that we want a sample check to give us.  Then we can calculate the rate of sampling that we need.  This is given in Figure~\ref{fig:samps} for confidence level $99.5\%$, and in Figure~\ref{fig:samps95} for confidence level $95\%$.

We can see from the chart how the level of sampling required changes with the prior probability $q$ of cheating,  particularly as $F$ increases.  For example we can see that if there is a $10\%$ prior probability of cheating, and we are concerned with whether $F = 100$ ballots have been changed, then we need to sample $3.0\%$ of the ballots, and find no cheating, to be able to conclude with $99.5\%$ confidence that this level of cheating has not occurred: in other words, the probability that there was no attack is $99.5\%$ given that this level of sampling finds no cheating.    However, we see that even with an extremely high prior probability of cheating of $50\%$, we still only need to sample $5.2\%$ of the ballots to achieve the same level of confidence that cheating has not occurred.

For a $95\%$ confidence level of no cheating on $F = 100$ and a prior probability of $10\%$ cheating, Figure~\ref{fig:samps95} shows that we only need to sample $0.7\%$ of the votes.

\begin{figure} 
\begin{center}
\includegraphics[width=\textwidth]{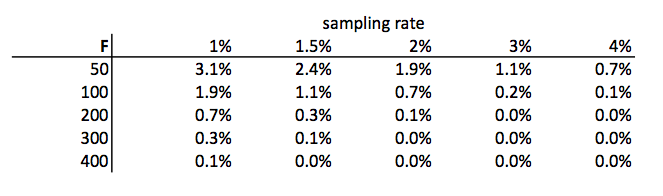}
\end{center}
\caption{Probability that that there was no cheating to the level of $F$ altered ballots for given sampling rates, with a prior probability of cheating of 5\%.} \label{fig:confchart}
\end{figure}

Conversely, Figure~\ref{fig:confchart} is appropriate for post-hoc analysis: given a value of $F$, and an observed level of sampling, the table gives the probability of there having been cheating to the level of $F$ altered ballots for a prior probability of 5\%.

Note that it is the voters themselves who decide to check the printing of the ballots, and who decide whether to confirm their votes on the WBB, so the rate of confirmation checks for ballot printing cannot be decided in advance.  However, mitigations such as additional print confirmation checks can be carried out by  ``mystery shoppers'' to achieve the required rate.

Reconsider the Bentleigh 2010 example with a winning margin of 522 votes.   Following our analysis, if there is no evidence of any attack then we might reasonably bound the prior probability of cheating at 10\%, (though this figure would require some careful consideration and seems at the high end).   Figure~\ref{fig:samps95} shows us that for a 95\% confidence level of no attack we need only sample 0.3\%  In fact a sampling rate of 1.2\% gives a confidence level of 99.5\%, as shown in Figure~\ref{fig:samps}.

%

\section{Probability that a sample passes the check given $F$ fraudulent ballots}
\label{app:BA}

Define:
\begin{itemize}
\item[$N$] The total population size (total number of ballots)
\item[$S$] The sample size: the number of ballots we will check
\item[$F$] The number of fraudulent (incorrectly constructed) ballots we are concerned about.
\item[$r$] The sample rate: sample size as a proportion of the total population size: $r = S/N$.
\end{itemize}
We'll also assume that $F \ll N$: that a very small proportion of the total ballots are fraudulent.

If there are $F$ fraudulent ballots out of $N$ it the following calculation determines the probability that a sample size of $S$ will not include any of the $F$ fraudulent ballots:
\begin{eqnarray*}
p & = & \Pi_{i = 0}^{S-1} ((N-F-i)/N-i)
\end{eqnarray*}
or
\begin{eqnarray*}
p & = & (\Pi_{i = 0}^{S-1} (N-F-i)) / (\Pi_{i = 0}^{S-1}(N-i))
\end{eqnarray*}
which simplifies to
\begin{eqnarray*}
p & = & (\Pi_{i = 0}^{F-1} (N-S-i)) / (\Pi_{i = 0}^{F-1}(N-i)) \\
  & = & \Pi_{i = 0}^{F-1} ((N-S-i) / (N-i)) \\
  & \approx & ((N-S)/N)^F \qquad \mbox{when $F \ll N$} \\
  & = & (1-r)^F
\end{eqnarray*}
Hence we can obtain the probability that the sample passes the check (i.e., discovers no fraud) given $F$ fraudulent ballots, will be approximately $(1-r)^F$.

\begin{figure} 
\begin{center}
\includegraphics[width=0.9\textwidth]{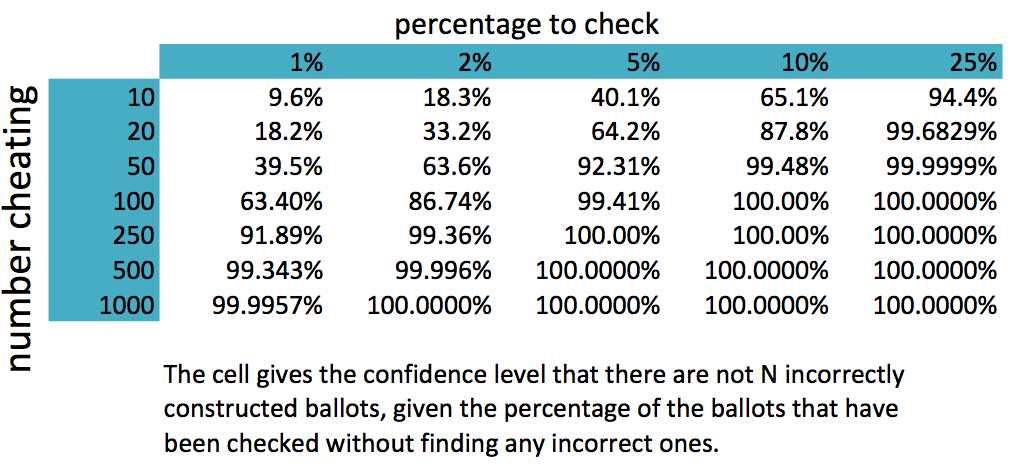}
\end{center}
\caption{Confidence levels} \label{fig:calcsex}
\end{figure}

A table of example confidence levels for sample sizes against number of ballots corrupted is given in Figure~\ref{fig:calcsex}.

\begin{figure} 
\begin{center}
\includegraphics[width=0.9\textwidth]{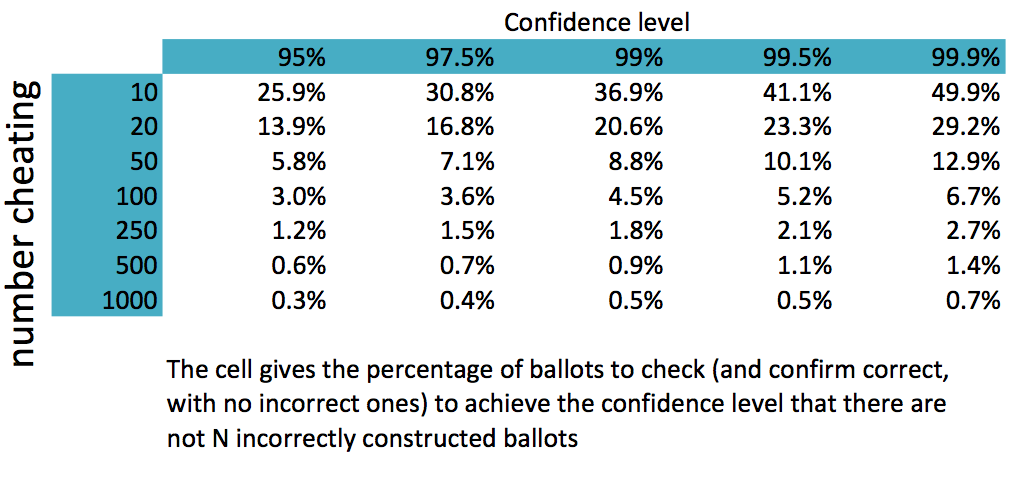}
\end{center}
\caption{Percentage sampled to achieve confidence levels} \label{fig:calcs2}
\end{figure}

A table of sample rates to achieve particular confidence levels against number of ballots corrupted is given in Figure~\ref{fig:calcs2}.  For example, to achieve a 99\% confidence that there are fewer than 1000 incorrect ballots, it is necessary to check 0.5\% of the total number of ballots.

\section{Bayesian Analysis}

\label{app:bayes}

Bayesian analysis is concerned with the way the probability of some occurrence $A$ is affected by particular observations $B$.   In our case we are interested in the probability that there are $F$ fraudulent ballots in the case where sampling does not identify any fraud.  Although we do not have a prior probability for fraudulent ballots, we we are still able to consider how sensitive the resulting probability is to possible prior probabilities.  

The calculations above give the probability that the sample passes the check given there are $F$ fraudulent ballots.  Define:
\begin{itemize}
\item[$A$]: there are $F$ fraudulent ballots
\item[$B$]: the sample passes the check
\item[$p(A)$]: the probability of $A$
\item[$p(\neg A)$]: the probability of $A$ being false
\item[$p(A | B)$]: the probability of $A$ given $B$
\end{itemize}

Bayes' theorem states that 
\begin{eqnarray}
p(A | B) & = & \frac{p(B | A) . p(A)}{p(B | A) . p(A) + p(B | \neg A) . p(\neg A)}  \label{bayes0}
\end{eqnarray}

For the worst case analysis we will assume that there are either no cheating ($0$ fraudulent ballots) or that there are exactly $F$ fraudulent ballots.  In other words, if there is an attack then the attacker chooses exactly the right number to change the result. 

We can therefore calculate $p = P(A | B)$, the probability we are interested,  given any particular value of $P(A)$, the prior probability that there has been an attack which altered $F$ ballots.

The formula in Line~\ref{bayes} above is Line~\ref{bayes0} with the following substitutions:
\begin{itemize}
\item $p(B | \neg A) = 1$ --- if no cheating then the probability that the sample passes the check must be $1$.;
\item $p(B | A) = (1-r)^F$ --- given earlier;
\item $p(A) = q$ --- prior (pre-sampling) probability of an attack.  Must be estimated;
\item $p(A | B) = p$ --- the probability we are interested in.
\end{itemize}


\end{document}